\documentclass[journal,twoside]{asmb}

\begin{document}
\title{On the Physical Layer Security Performance over RIS-aided Dual-hop RF-UOWC Mixed Network}

\author[1]{T Hossain}
\author[2]{S. Shabab}
\author[3]{A. S. M. Badrudduza}
\author[4]{M. K. Kundu}
\author[5]{I S Ansari }

\affil[1,3]{Department of Electronics \& Telecommunication Engineering, Rajshahi University of Engineering \& Technology (RUET), Rajshahi-6204, Bangladesh}
\affil[2]{Department of Electrical \& Electronic Engineering, RUET}
\affil[4]{Department of Electrical \& Computer Engineering, RUET}
\affil[5]{James Watt School of Engineering, University of Glasgow, Glasgow G12 8QQ, United Kingdom}

\twocolumn[
\begin{@twocolumnfalse}
\maketitle
\begin{abstract}
\section*{Abstract}

Since security has been one of the crucial issues for high-yield communications such as $5$G and $6$G, the researchers continuously come up with newer techniques to enhance the security and performance of these progressive wireless communications.
Reconfigurable intelligent surface (RIS) is one of those techniques that artificially rearrange and optimize the propagation environment of electromagnetic waves to improve both spectrum and energy efficiency of wireless networks. Besides, in underwater communication, underwater optical wireless communication (UOWC) is a better alternative/replacement for conventional acoustic and radio frequency (RF) technologies. Hence, mixed RIS-aided RF-UOWC can be treated as a promising technology for future wireless networks. This work focuses on the secrecy performance of mixed dual-hop RIS-aided RF-UOWC networks under the intercepting effort of a probable eavesdropper. The RF link operates under generalized Gamma fading distribution; likewise, the UOWC link experiences the mixture exponential generalized Gamma distribution. The secrecy analysis subsumes the derivations of closed-form expressions for average secrecy capacity, exact and lower bound of secrecy outage probability, and strictly positive secrecy capacity, all in terms of Meijer’s $G$ functions. Capitalizing on these derivations, the effects of heterodyne and intensity modulation/direct detection systems, underwater turbulence resulting from air bubble levels, temperature gradients, and salinity gradients, are measured. Unlike conventional models that merely deal with thermally uniform scenarios, this proposed model is likely to be unique in terms of dealing with secrecy analysis of a temperature gradient RIS-aided RF-UOWC network. Lastly, the derivations are validated via Monte-Carlo simulations.
\end{abstract}
\begin{IEEEkeywords}
\section*{Keywords} 

Mixture exponential generalized Gamma, physical layer security, reconfigurable intelligent surface, underwater optical wireless communications.
\end{IEEEkeywords}
\end{@twocolumnfalse}
]

\section{Introduction}

\subsection{Background and Literature Study}
The fifth generation ($5$G) wireless communications, performing with inflated data rates, highly capable base stations, and low latency, plays a pivotal role to provide robust development of billions of devices \cite{alsharif2017evolution}. This technological blessing helps to fulfil the essential need, complying with larger data traffic of mobile communications in every sphere. To adapt with the current situation, researchers offered and installed different solutions such as mmWaves, massive multiple-input and multiple-output (m-MIMO), small cells technique, reconfigurable intelligence surface (RIS), non-orthogonal multiple access (NOMA), ultra-dense heterogeneous networks, etc., incorporating both free space and underwater optical wireless communication (UOWC) networks that ensure expected performance according to the global traffic demand \cite{alsharif2017evolution, ibrahim2021enhancing}.

Reconfigurable intelligent surfaces (RIS) are drawing enormous attention of the research community as an efficient mechanism for future secure wireless communication systems. This technology has significantly enhanced the communication quality by making the propagation environment programmable \cite{li2021performance}. A good number of outstanding features of RIS such as passive beamforming, and customization of propagation medium by means of adjustable phase and amplitude, have created a new era in wireless communications \cite{makarfi2020physical}. In fact, the RIS is made of low cost and energy efficient reflecting reconfigurable elements that can reflect the incoming signals inducing modifiable phase shifts, which are further adjusted and then aligned towards the desired destination. Some promising applications are installed using RIS techniques, e.g., enhancing the m-MIMO systems, maximizing signal-to-noise ratio (SNR), promoting the signal coverage, optimising the beamforming of multi-user channels \cite{li2021performance}, etc.

In \cite{huang2019reconfigurable}, a realistic RIS-based system was presented where the authors obtained higher energy efficiency in comparison with the simple amplify-and-forward (AF) multi-antenna systems. An RIS-assisted multiple-input single-output (MISO) network was studied in \cite{kammoun2020asymptotic} where the author found that RIS with small number of passive elements outperforms half-duplex relaying but RIS with large number of passive elements is needed to outperform the full-duplex relaying. In \cite{hou2020reconfigurable}, RIS-aided NOMA network was developed that showed superior performance relative to other orthogonal networks regarding spectral efficiency (SE), energy efficiency (EE), outage probability, and ergodic capacity. RIS-assisted optical wireless communication networks were also gaining popularity where the RIS is utilized along with the optical networks in order to promote the system performance with an improved coverage \cite{yang2020mixed, li2021performance}. RIS was used as a promising solution for free-space optical (FSO) communications to overcome atmospheric turbulence problem and pointing error issues in \cite{6952039,6777774,yang2020free}. The performance of RIS-empowered FSO system was demonstrated unifying Fisher–Snedecor ($\mathcal{F}$), Gamma-Gamma (GG), and M\'alaga ($\mathcal{M}$) distributions in \cite{chapala2021unified}. RIS-assisted decode-and-forward (DF) relaying UOWC system was investigated in \cite{odeyemi2020performance} where the authors showed the RIS element numbers, detection techniques, and UOWC optical turbulence, all affect overall system outage probability and average bit error rate (ABER). The effect of interference in a FSO-radio frequency (RF) system assisted by RIS is observed in \cite{sikri2021reconfigurable} where the authors showed that the performance of this scenario can be improved by increasing FSO apertures.

UOWC is playing a leading key role in real-time sub-sea disciplines such as oceanography, offshore communications, sea climate observations, military, industry, and various sea-related engineering research activities \cite{zeng2016survey}. In modern times, four types of major communications are included in UOWC: Acoustic wave (AW) communications, FSO communications, RF communications, and magnetic induction (MI) communications \cite{singh2019underwater}. The transmitting signal of UOWC is generated from conical beams of laser or LED lights and $400$-$550$ nm spectrum band at blue or green portion of seawater \cite{zedini2019unified}. This technology might be a promising solution of high data-rate, bandwidth, transmission security, seawater attenuation, and it also hugely impacts low cost LEDs, digital signal processing, digital multiplexing, and digital communications \cite{zeng2016survey}. The widely used UOWC model is the mixture exponential generalized Gamma (mEGG) model since this model provides a good fit for all the experimental data and at the same time it possesses good analytical properties. The expressions of ABER and outage probability of RF-UOWC system were analyzed both in closed-forms and asymptotically in \cite{lei2020performance}. The RF link was further generalized in \cite{badrudduza2021security}. In \cite{yang2021performance}, a dual-hop FSO-UOWC system was shown where authors noticed significant impacts of the pointing errors, water bubbles, and temperature gradients on the system performance.

Since the people are relying heavily on the wireless networks for exchanging private information, the capability of sharing secret information in the presence of eavesdroppers is extremely important. In this respect, physical layer security (PLS) approach proposed by \cite{wyner1975wire} is growing popularity that eliminates the pitfalls of secret key management in the classical cryptographic methods \cite{badrudduza2020enhancing}. RIS can be a promising solution for PLS because of two reasons: (a) RIS makes the main channels stronger relative to the eavesdropper channels as a result of co-phasing all the reflected signals with the received transmitted signal via the direct link, and (b) RIS suppresses the eavesdroppers' channel if the eavesdropper receives the reflected signal with opposite phase with respect to that of the transmitter. In \cite{wang2020intelligent}, a beamforming and jamming technique were presented jointly to analyze the secrecy performance without any eavesdropper's channel state information (CSI). The impacts of colluding and non-colluding eavesdroppers assuming a discrete phase shift at the RIS was shown in \cite{xu2020ergodic} in terms of ergodic secrecy rate. The PLS in vehicular network was described in \cite{makarfi2020physical}, where the transmission is dependent on RIS, source power, eavesdroppers' position, and distance. Relative to the RF links that can easily be wiretapped, the optical link exhibits more secure characteristics \cite{sarker2021effects, islam2021impact, lou2021secrecy}. A DF related RF-UOWC network was shown in \cite{lou2021secrecy}, where the secrecy outage performance was exhibited using EGG model for the UOWC link, which is further generalized using mEGG model in \cite{badrudduza2021security}. The same UOWC model was also utilized by \cite{ibrahim2021enhancing}, where the authors demonstrated that transmit antenna selection / maximal ratio combining (TAS / MRC) diversity can reduce the detrimental impacts of underwater turbulence (UWT). An innovative approach was used in the case of a highly secure underwater communication scenario considering cross layering and context-centric networking with state-of-the-art counter measurement in \cite{lal2016secure}. In \cite{illi2018dual}, the average secrecy capacity (ASC) and secrecy intercept probability (SIP) performances of RF-UOWC system were shown. SIP was also analyzed in \cite{illi2018secrecy} with amplify-and-forward (AF) relaying making use of the MRC technique at the relay.

\subsection{Motivation and Contributions}
Over the recent times, researchers have proposed a few number of security models over UOWC channels and most of the proposed models are related to the relay assisted RF-UOWC mixed systems. Specifically, there exists no analysis in the literature on the security domain of RIS-assisted UOWC model. Hence, secrecy performance analysis over RIS-aided RF-UOWC system, to the very best of authors’ knowledge based on the open literature, can be seen as a novel application scenario. In this regard, we propose a RIS-aided RF-UOWC mixed network wherein the RF link follows GG distribution and the UOWC link undergoes mEGG distribution. A relay node, which exists in between the system source and the user as the destination, performs three tasks accordingly: Receiving RF signals from the source, transforming the signal to an optical structure, and passing the signal to the targeted user through the UOWC link. 
when the relay accepts the RF signal, an eavesdropper simultaneously receives the  transmitted  data  from the same  RF  link.  
As compared to the Rician fading channel in \cite{salhab2021accurate}, the offered Nakagami-$m$ model offers a more versatile scenario and several classical RIS models (e.g. Rayleigh in \cite{yang2020accurate}) can be shown as a special case of this model for various values of its shape parameter. On the other hand, the considered mEGG model has the capability to adopt all of the pragmatically occurred physical transformations such as UWT, air bubbles, temperature gradient, and water salinity \cite{zedini2019unified}, \cite{illi2018dual}. Besides, this model also serves as one of the best models facilitating a lot of analytical tractability in deriving expressions of secrecy performance metrics. For instance, the widely used EGG model can be obtained as a special case of mEGG model \cite{illi2019physical}. The major contributions of this research are mentioned as follows:

\begin{enumerate}

\item At first, we drive the cumulative density function (CDF) of the dual-hop SNR for the RIS-aided RF-UOWC network. Note that the proposed RIS-aided network over (Nakagami-$m$)-mEGG model represents a unique application scenario, and hence the derived CDF is also unique. As compared to \cite{li2021performance}, we consider both RIS-aided RF links i.e. both source-to-RIS and RIS-to-relay follow Nakagami-$m$ distribution whereas \cite{li2021performance} assumed RIS-to-relay link only to follow the Rayleigh model that is a mere special case of the proposed model.

\item The secrecy performance of the proposed network is demonstrated with respect to the average secrecy capacity (ASC), exact and lower bound of secrecy outage probability (SOP), and strictly positive secrecy capacity (SPSC), etc., expressions in closed-form and further corroborated via Monte-Carlo (MC) simulations. To the best of authors' knowledge based on the open literature, these expressions are novel and generalized, and can be utilized to unify versatile classical existing models.

\item Capitalizing on the derived expressions, noticeable impacts of air bubbles and temperature gradients based UWTs for both salty and fresh waters along with the RIS-aided RF channel parameters are also demonstrated. At last, we analyse effects of two types of detection techniques i.e. intensity modulation/direct detection (IM/DD), and heterodyne detection (HD) techniques wherein the HD technique seems more efficient in secrecy improvement relative to the IM/DD technique.

\end{enumerate}

\begin{figure*}[!ht]
\vspace{0mm}
    \centerline{\includegraphics[width=0.9\textwidth,angle =0]{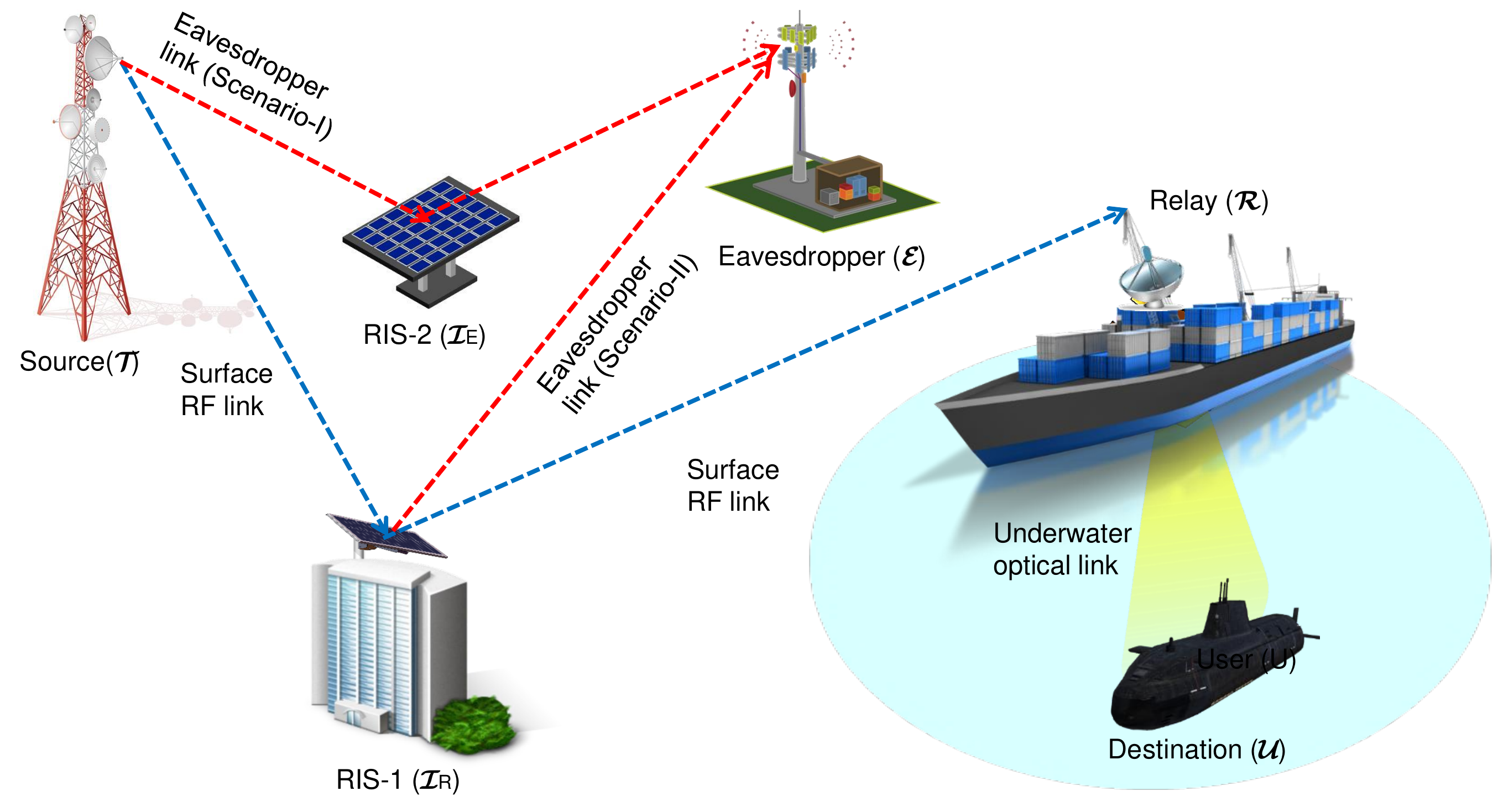}}
        \vspace{0mm}
    \caption{System model of a combined RIS-aided dual-hop RF-UOWC system with source ($\mathcal{T}$), relay ($\mathcal{R}$), user ($\mathcal{U}$), and eavesdropper ($\mathcal{E}$).}
    \label{fig:1}
\end{figure*}
\subsection{Organization}

The paper's subsequent parts are categorized as follows. The proposed system, a mixed RIS-aided RF-UOWC model, and the PDFs and CDFs for each individual links are represented in Section II. Section III mathematically demonstrates the derivation of the expressions for ASC, exact SOP, lower bound of SOP, and SPSC. Numerical and simulation results are explained in Section IV. Finally, Section V summarizes the output of our works.

\section{System Model and Problem Formulation}
\label{s1}
As demonstrated in Fig. \ref{fig:1}, we present the system model of a generic dual hop RIS-aided RF-UOWC system consisting of a source $\mathcal{T}$, an intermediate relay ($\mathcal{R}$), two RISs,  $\mathcal{I}_{R}$ and $\mathcal{I}_{E}$, and a destination user $\mathcal{U}$ that is stationed underneath the water surface. The distance between $\mathcal{T}$ and $\mathcal{R}$ is assumed very large, and due to the environmental structures between these two nodes, $\mathcal{T}$ and $\mathcal{R}$ are not connected via any direct link. Hence, the transmitted signal from $\mathcal{T}$ is first sent to $\mathcal{I}_{R}$ that in turn reflects it to $\mathcal{R}$. The RIS is assumed to be capable of obtaining the CSI of the $\mathcal{T}-\mathcal{R}$ link and this CSI is further utilized to maximize the received SNR at $\mathcal{R}$. Mainly, we can adjust the RIS induced phases to maximise the received SNR at $\mathcal{R}$ by means of necessary phase cancellations and appropriate alignment of reflected signals from RIS \cite{basar2019wireless}. The private RIS-aided communication between $\mathcal{T}$ and $\mathcal{U}$ through $\mathcal{R}$ is overheard by a fortuitous eavesdropper $\mathcal{E}$ that attempts to wiretap using similar RIS aided $\mathcal{T}-\mathcal{I}_{E}-\mathcal{E}$ link. We consider following two eavesdropping scenarios.

\begin{itemize}

\item In \textit{Scenario-I}, $\mathcal{R}$ and $\mathcal{E}$ utilize two different RISs, i.e., $\mathcal{I}_{R}$ and $\mathcal{I}_{E}$, respectively. Hence, the received incoming reflected signals at $\mathcal{R}$ and $\mathcal{E}$ are also different.

\item The \textit{Scenario-II} assumes there exists only one RIS and both $\mathcal{R}$ and $\mathcal{E}$ receive the same reflected signal from $\mathcal{I}_{R}$. It is noteworthy that \textit{Scenario-II} can be achieved as special case of \textit{Scenario-I}.

\end{itemize}
\noindent
Similar to \cite{yang2020secrecy}, the RIS is assumed unaware of the eavesdropper's CSI and hence, the considered model represents a passive eavesdropping scenario. Recently, this particular framework is more appreciable because of performing better than conventional schemes and also satisfying all the requirements demanded by several emerging applications such as unmanned aerial vehicles (UAVs) \cite{li2020reconfigurable}, large intelligent surfaces/antennas (LISA) technology \cite{liang2019large}, undersea vehicles \cite{zeng2016survey}, industrial IoT \cite{vitturi2019industrial}, etc. Herein, $\mathcal{T}$ and $\mathcal{E}$ are provided with a single antenna while $\mathcal{R}$ is furnished with a single transmit aperture along with a single receive antenna. To receive the optical wave, $\mathcal{U}$ has a single photo detector while $\mathcal{I}_{R}$ and $\mathcal{I}_{E}$ have $S_{1}$ and $S_{2}$ number of reflecting elements, respectively. As $\mathcal{T}$ is not directly connected to $\mathcal{U}$, we consider overall transmission to occur in two hops. In the first hop, the system model considers $\mathcal{T}$ to transmit signals aided with a RIS $\mathcal{I}_{R}$ via the $\mathcal{T}-\mathcal{I}_{R}-\mathcal{R}$ link to reach $\mathcal{R}$ in the presence of $\mathcal{E}$ that uses the same ($\mathcal{I}_{R}$) / another ($\mathcal{I}_{E}$) RIS-aided RF link ($\mathcal{T}-\mathcal{I}_{E}-\mathcal{E}$) to conduct surveillance. The surface RF networks using $\mathcal{T}-\mathcal{I}_{R}-\mathcal{R}$, $\mathcal{T}-\mathcal{I}_{R}-\mathcal{E}$, and $\mathcal{T}-\mathcal{I}_{E}-\mathcal{E}$ links follow the Nakagami-$m$ fading distribution. The relay is utilized to transform the received RF signal into respective optical form and then redirects it to the underwater user $\mathcal{U}$. This underwater network utilizing $\mathcal{R}-\mathcal{U}$ UOWC link experiences mEGG distribution.

\subsection{SNRs of Individual Links}
For \textit{Scenario-I}, let us denote $h_{i_{1}}$ ($i_{1}=1, 2, \ldots, S_{1}$) and $m_{i_{2}}$ ($i_{2}=1, 2, \ldots, S_{2}$) as the first hop channel gains of the $\mathcal{T}-\mathcal{I}_{R}-\mathcal{R}$ and $\mathcal{T}-\mathcal{I}_{E}-\mathcal{E}$ links, respectively. Similarly, the second hop channel gains of the $\mathcal{T}-\mathcal{I}_{R}-\mathcal{R}$ and $\mathcal{T}-\mathcal{I}_{E}-\mathcal{E}$ links are denoted by $g_{i_{1}}$ and $n_{i_{2}}$, respectively. Hence, the received signals at $\mathcal{R}$ and $\mathcal{E}$ are, respectively, given by
\begin{align}
\label{r1}
y_{t,r}&=\left[\sum_{i_{1}=1}^{S_{1}}
h_{i_{1}}e^{j\Theta_{i_{1}}}g_{i_{1}}\right]x+z_{1},
\\
\label{r2}
y_{t,e}&=\left[\sum_{i_{2}=1}^{S_{2}}m_{i_{2}}e^{j\theta_{i_{2}}}n_{i_{2}}\right]x+z_{2},
\end{align} 
For these concerned channels, we have $h_{i_{1}}=\alpha_{i_{1}}e^{j\varrho_{i_{1}}}$, $m_{i_{2}}=\delta_{i_{2}}e^{j\varrho_{i_{2}}}$,  $g_{i_{1}}=\beta_{i_{1}}e^{j\vartheta_{i_{1}}}$, and $n_{i_{2}}=\eta_{i_{2}}e^{j\vartheta_{i_{2}}}$, where $\alpha_{i_{1}}$, $\beta_{i_{1}}$, $\delta_{i_{2}}$, and $\eta_{i_{2}}$ are the Nakagami-$m$ distributed random variables (RVs), $\varrho_{i_{1}}$, $\varrho_{i_{2}}$, $\vartheta_{i_{1}}$, and $\vartheta_{i_{2}}$ are the phases of the channel gains, $\Theta_{i_{1}}$, $\theta_{i_{2}}$ denote the adjustable phases produced by the $i_{1}$-th and $i_{2}$-th reflecting element of the RISs, $x$ symbolises the transmitted data symbol from $\mathcal{T}$ with power $T_{s}$, and $z_{1}\sim\mathcal{\widetilde{N}}(0,N_{r})$, $z_{2}\sim\mathcal{\widetilde{N}}(0,N_{e})$ are the additive white Gaussian noise (AWGN) samples with $N_r$, $N_e$ representing the noise power of the corresponding channels. In matrix form, \eqref{r1} and \eqref{r2} are written as
\begin{align}
\label{r3}
y_{t,r}&=\textbf{g}^{T}\Theta\,\textbf{h}\,x+z_{1},
\\
\label{r4}
y_{t,e}&=\textbf{n}^{T}\theta\,\textbf{m}\,x+z_{2},
\end{align} 
where $\textbf{h}=[h_{1}\,h_{2}\,\ldots\,h_{S_{1}}]^{T}$, $\textbf{g}=[g_{1}\,g_{2}\,\ldots\,g_{S_{1}}]^{T}$, $\textbf{m}=[m_{1}\,m_{2}\,\ldots\,m_{S_{2}}]^{T}$ and $\textbf{n}=[n_{1}\,n_{2}\,\ldots\,n_{S_{2}}]^{T}$ denote the channel coefficient vectors, and $\Theta=\text{diag}([e^{j\Theta_{1}}\,e^{j\Theta_{2}}\,\ldots\, e^{j\Theta_{S_{1}}}])$ and $\theta=\text{diag}([e^{j\theta_{1}}\,e^{j\theta_{2}}\,\ldots\, e^{j\theta_{S_{2}}}])$ are the diagonal matrices that incorporates the phase shifts applied by the RIS elements. We can now represent the instantaneous SNRs at $\mathcal{R}$ and $\mathcal{E}$ as
\begin{align}
\gamma_{\mathcal{R},I,\Theta}&=\frac{\left[\sum_{i_{1}=0}^{S_{1}}\alpha_{i_{1}}\beta_{i_{1}}e^{j\left(\Theta_{i_{1}}-\varrho_{i_{1}}-\vartheta_{i_{1}}\right)}\right]^{2} T_{s}}{N_{r}},
\\
\gamma_{\mathcal{E},I,\theta}&=\frac{\left[\sum_{i_{2}=0}^{S_{2}}\delta_{i_{2}}\eta_{i_{2}}e^{j\left(\theta_{i_{2}}-\varrho_{i_{2}}-\vartheta_{i_{2}}\right)}\right]^{2} T_{s}}{N_{e}}.
\end{align} 
It is noteworthy the optimal choice of $\Theta_{i_{1}}$ and $\theta_{i_{2}}$ that maximizes the instantaneous SNR is $\Theta_{i_{1}}=\varrho_{i_{1}}+\vartheta_{i_{1}}$ and $\theta_{i_{2}}=\varrho_{i_{2}}+\vartheta_{i_{2}}$, respectively. Hence, the maximized SNRs at $\mathcal{R}$ and $\mathcal{E}$ are expressed as
\begin{align}
\gamma_{\mathcal{R},I}&=\frac{\left(\sum_{i_{1}=0}^{S_{1}}\alpha_{i_{1}}\beta_{i_{1}}\right)^{2}T_{s}}{N_{r}}=\left(\sum_{i_{1}=0}^{S_{1}}\alpha_{i_{1}}\beta_{i_{1}}\right)^{2} \gamma_{m_{1}},
\\
\gamma_{\mathcal{E},I}&=\frac{\left(\sum_{i_{2}=0}^{S_{2}}\delta_{i_{2}}\eta_{i_{2}}\right)^{2}T_{s}}{N_{e}}=\left(\sum_{i_{2}=0}^{S_{2}}\delta_{i_{2}}\eta_{i_{2}}\right)^{2} \gamma_{m_{2}},
\end{align} 
where $\gamma_{{m}_1}$ denotes the average SNR of the $\mathcal{T}-\mathcal{I}_{R}-\mathcal{R}$ link and $\gamma_{{m}_2}$ indicates the average SNR of $\mathcal{T}-\mathcal{I}_{E}-\mathcal{E}$ link.

\quad

\noindent
\textbf{Special Case:}

The received signals corresponding to \textit{Scenario-II} are expressed similar to \eqref{r1} and \eqref{r2} while considering $h_{i_{1}}=m_{i_{2}}=f_{i_{3}}=\xi_{i_{3}}e^{j\Xi_{i_{3}}}$ and $S_{1}=S_{2}=S_{3}$ with adjustable phase $\Xi_{i_{3}}$ induced by $\mathcal{I}_{R}$. The corresponding maximised SNRs are obtained by letting RVs $\alpha_{i_{1}}=\delta_{i_{2}}=\xi_{i_{3}}$ as
\begin{align}
\gamma_{\mathcal{R},II}&=\frac{\left(\sum_{i_{3}=0}^{S_{3}}\xi_{i_{3}}\beta_{i_{3}}\right)^{2}T_{s}}{N_{r}}=\left(\sum_{i_{3}=0}^{S_{3}}\xi_{i_{3}}\beta_{i_{3}}\right)^{2}\gamma_{m_{1}},
\\
\gamma_{\mathcal{E},II}&=\frac{\left(\sum_{i_{3}=0}^{S_{3}}\xi_{i_{3}}\eta_{i_{3}}\right)^{2}T_{s}}{N_{e}}=\left(\sum_{i_{3}=0}^{S_{3}}\xi_{i_{3}}\eta_{i_{3}}\right)^{2}\gamma_{m_{2}},
\end{align}

where $\beta_{i_3}$ and $\eta_{i_3}$ are Nakagami-$m$ distributed RVs. Note that the SNRs $\gamma_{\mathcal{R},I}$ and $\gamma_{\mathcal{E},I}$ are referred to as $\gamma_{\mathcal{R}}$ and $\gamma_{\mathcal{E}}$ in the remaining manuscript, respectively.

Denoting the direct channel gain between $\mathcal{R}$ and $\mathcal{U}$ as $l_{r,u}$ $\in$ $\C^{1\times1}$, the expression for received signals at $\mathcal{U}$ is expressed as
\begin{align}
y_{r,u}&=l_{r,u}\,y_{t,r}+z_{u},
\end{align}
where $z_{u}\sim\mathcal{\widetilde{N}}(0,N_{u})$ exemplifies optical noises imposed at $\mathcal{U}$ and $N_{u}$ is noise power at the destination. The corresponding SNR is given as
\begin{align}
\gamma_{\mathcal{U}}&=\frac{T_{r}}{N_{u}}\|l_{r,u}\|^{2}=\gamma_{m_{u}}\|l_{r,u}\|^{2}.
\end{align}
Note that the gain related to electrical-to-optical conversion while relaying is incorporated altogether in the $\mathcal{R}-\mathcal{U}$ link gain. The received SNR of the combined RIS-aided RF-UOWC system while utilizing AF variable gain relaying scheme is given as \cite[Eq.~(28)]{juel2021secrecy}
\begin{align}
\gamma_{eq}&=\frac{\gamma_{\mathcal{R}}\,\gamma_{\mathcal{U}}}{\gamma_{\mathcal{R}}+\gamma_{\mathcal{U}}+1} \cong min\left\{\gamma_{\mathcal{R}}, \gamma_{\mathcal{U}}\right\}.
\end{align} 

\subsection{PDF and CDF of $\gamma_{\mathcal{R}}$}

The PDF of $\gamma_{\mathcal{R}}$ (\textit{Scenario-I}) is expressed as \cite[Eq.~(2)]{samuh2020performance}
\begin{align}
    \label{eq1}
    f_{\gamma_{\mathcal{R}}}(\gamma)=K_{{r}_{1}}\gamma^{K_{{r}_{2}}} e^{-K_{{r}_{3}}\gamma^{\frac{1}{2}}},
\end{align} 
where $K_{{r}_{1}}=\frac{\gamma_{m_{1}}^{-\frac{a_{r}+1}{2}}}{(2b_{r})^{a_{r}+1}\,\Gamma(a_{r}+1)}$, $K_{r_{2}}=\frac{a_{r}-1}{2}$, $K_{{r}_{3}}=\frac{\sqrt\frac{1}{\gamma_{{m}_{1}}}}{b_{r}}$,
$a_{r}=\frac{m_{r_{1}}m_{r_{2}}S_{1}\Gamma\left(m_{r_{1}}\right)^{2}\Gamma\left(m_{r_{2}}\right)^{2}}{m_{r_{1}}m_{r_{2}}\Gamma\left(m_{r_{1}}\right)^{2}\Gamma\left(m_{r_{2}}\right)^{2}-\Gamma\left(m_{r_{1}}+\frac{1}{2}\right)^{2}\Gamma\left(m_{r_{2}}+\frac{1}{2}\right)^{2}}-S_{1}-1$, and $b_{r}=\frac{m_{r_{1}}m_{r_{2}}\Gamma\left(m_{r_{1}}\right)^{2}\Gamma\left(m_{r_{2}}\right)^{2}-\Gamma\left(m_{r_{1}}+\frac{1}{2}\right)^{2}\Gamma\left(m_{r_{2}}+\frac{1}{2}\right)^{2}}{\sqrt{\frac{m_{r_{1}}}{\Omega_{r_{1}}}}\Gamma\left(m_{r_{1}}\right)\Gamma\left(m_{r_{1}}+\frac{1}{2}\right)\sqrt{\frac{m_{r_{2}}}{\Omega_{r_{2}}}}\Gamma\left(m_{r_{2}}\right)\Gamma\left(m_{r_{2}}+\frac{1}{2}\right)}$. For the first hop of $\mathcal{T}-\mathcal{I}_{R}-\mathcal{R}$ link, $m_{r_{1}}$ and $\Omega_{r_{1}}$ denote the fading parameter and scale parameter or mean power (second moment), respectively, and for the second hop they are denoted by $m_{r_{2}}$ and $\Omega_{r_{2}}$, respectively, where $\Gamma(.)$ designates the Gamma operator. The CDF of $\gamma_{\mathcal{R}}$ is defined as
\begin{align}
\label{eq2}
F_{\gamma_{\mathcal{R}}}(\gamma)=\int_{0}^{\gamma}f_{\gamma_{\mathcal{R}}}(\gamma)d\gamma.
\end{align} 
Substituting \eqref{eq1} in \eqref{eq2} and exploiting \cite[Eq.~(3.381.8)]{gradshteyn2014table}, $F_{\gamma_{\mathcal{R}}}(\gamma)$ is followed as
\begin{align}
    \label{eq3}
    F_{\gamma_{\mathcal{R}}}(\gamma)=\frac{\Upsilon\left(a_{r}+1,\frac{\sqrt\frac{\gamma}{\gamma_{{m}_{1}}}}{b_{r}}\right)}{\Gamma(a_{r}+1)},
\end{align} 
where $\Upsilon(.,.)$ is the lower incomplete Gamma function \cite[Eq.~(8.350.1)]{gradshteyn2014table}. Exploiting identity \cite[Eq.~(8.354.1)]{gradshteyn2014table}, \eqref{eq3} becomes
\begin{align}
\label{eq4}
F_{\gamma_{\mathcal{R}}}(\gamma)=\sum_{n_{1}=0}^{\infty}K_{r_{4}}\gamma^{\frac{\upsilon_{1}}{2}},
\end{align} 
where $K_{r_{4}}=\frac{(-1)^{n_{1}}K_{{r}_{3}}^{\upsilon_{1}}}{n_{1}!\,\upsilon_{1}}$ and $\upsilon_{1}=a_{r}+n_{1}+1$.

\subsection{PDF and CDF of $\gamma_{\mathcal{U}}$}

The UOWC link follows mEGG distribution. The PDF of $\gamma_{\mathcal{U}}$ considering both HD and IM/DD techniques is as \cite{zedini2019unified}
\begin{align}
\label{eq5}
f_{\gamma_{\mathcal{U}}}(\gamma)=\sum_{i=1}^{2}{\Upsilon}_{i}\,\gamma^{-1}\,G_{0,1}^{1,0}\left[{\Psi}_{i}\gamma^{{\Lambda}_{i}}\biggl|
\begin{array}{c}
 - \\
{\chi}_{i} \\
\end{array}
\right],
\end{align} 
where ${\Psi}_{1}=\frac{1}{\sigma\,\eta_{s}^{\frac{1}{s}}}$, ${\Lambda}_{1}=\frac{1}{s}$, ${\chi}_{1}=1$, ${\Upsilon}_{1}=\frac{\lambda}{s}$, ${\Psi}_{2}=\frac{1}{q^{r}\eta_{s}^{\frac{r}{s}}}$, ${\Lambda}_{2}=\frac{r}{s}$, ${\chi}_{2}=p$, ${\Upsilon}_{2}=\frac{r\,(1-\lambda)}{s\,\Gamma(p)}$, $0<\lambda<1$ signifies the mixture weight, $\sigma$ designates the exponential distribution parameter, and $\eta_{s}$ denotes the electrical SNR of UOWC link. Here, $p$, $q$, and $r$ indicate the GG distribution parameters and $G_{c,d}^{a,b}[.|.]$ is the Meijer's $G$ function \cite{gradshteyn2014table}. Detection technique is indicated by $s$ that formulates HD technique for $s=1$ and IM/DD technique for $s=2$ conditions. Electrical SNR for HD technique and IM/DD technique are specified as $\eta_{1}=\gamma_{m_{u}}$ and $\eta_{2}=\frac{\gamma_{m_{u}}}{2\,\lambda\,\sigma^{2}+b^{2}(1-\lambda)\frac{\Gamma\left(p+\frac{2}{r}\right)}{\Gamma\left(p\right)}}$, respectively.

To speculate the effects of several bubble levels and temperature gradients on turbulence scenarios and water salinity, values of $\lambda$, $\sigma$, $p$, $q$, and $r$ went under trial run in \cite{zedini2019unified}. Table $2$ of \cite{badrudduza2021security} shows the increase in air bubble levels denoted by $h$ with temperature gradient denoted by $l$ that produces weak, average, and strong turbulence conditions, respectively. On the other hand, water salinity considering thermally uniform UOWC network is observed in Table $3$ of \cite{badrudduza2021security} to present different turbulence scenarios due to several air bubble levels for both fresh and salty waters. Hence, mEGG model helps to carry performance analysis under different turbulence scenarios in both thermal gradient and thermally uniform UOWC systems that makes this model more acceptable in the research field. Following the similar process as in \eqref{eq2}, the CDF for the SNR of UOWC link is expressed as \cite{zedini2019unified}
\begin{align}\label{eq6}
F_{\gamma_{\mathcal{U}}}(\gamma)=\sum_{i=1}^{2}\zeta_{i}\,G_{1,2}^{1,1}\left[{\Psi}_{i}\gamma^{{\Lambda}_{i}}\biggl|
\begin{array}{c}
 1 \\
\mathcal{\chi}_{i},0 \\
\end{array}
\right],
\end{align} 
where ${\zeta}_{1}=\lambda$ and ${\zeta}_{2}=\frac{1-\lambda}{\Gamma\left(p\right)}$.

\subsection{PDF and CDF of $\gamma_{\mathcal{E}}$}

Presuming $\mathcal{T}-\mathcal{I}_{E}-\mathcal{E}$ link (\textit{Scenario-I}) also considers Nakagami-$m$ distribution, the PDF and CDF of $\gamma_{\mathcal{E}}$ are expressed as \cite[Eq.~(2) and (3)]{lei2015physical}
\begin{align}
\label{eq7}
f_{\gamma_{\mathcal{E}}}(\gamma)={K}_{{e}_{1}}\gamma^{{K}_{{e}_{2}}}e^{-{K}_{{e}_{3}}\gamma^{\frac{1}{2}}},
\end{align}
and
\begin{align}
\label{eq8}
F_{\gamma_{\mathcal{E}}}(\gamma)=\sum_{n_{2}=0}^{\infty}K_{{e}_{4}}\gamma^{\frac{\upsilon_{2}}{2}},
\end{align} 
where $K_{{e}_{1}}$=$\frac{\gamma_{m_{2}}^{-\frac{a_{e}+1}{2}}}{\left(2b_{e}\right)^{a_{e}+1}\,\Gamma(a_{e}+1)}$, $K_{e_{2}}$=$\frac{a_{e}-1}{2}$, $K_{{e}_{3}}=\frac{\sqrt\frac{1}{\gamma_{{m}_{2}}}}{b_{2}}$, 
$a_{e}=\frac{m_{e_{1}}m_{e_{2}}S_{2}\Gamma\left(m_{e_{1}}\right)^{2}\Gamma\left(m_{e_{2}}\right)^{2}}{m_{e_{1}}m_{e_{2}}\Gamma\left(m_{e_{1}}\right)^{2}\Gamma\left(m_{e_{2}}\right)^{2}-\Gamma\left(m_{e_{1}}+\frac{1}{2}\right)^{2}\Gamma\left(m_{e_{2}}+\frac{1}{2}\right)^{2}}-S_{2}-1$, $b_{e}=\frac{m_{e_{1}}m_{e_{2}}\Gamma\left(m_{e_{1}}\right)^{2}\Gamma\left(m_{e_{2}}\right)^{2}-\Gamma\left(m_{e_{1}}+\frac{1}{2}\right)^{2}\Gamma\left(m_{e_{2}}+\frac{1}{2}\right)^{2}}{\sqrt{\frac{m_{e_{1}}}{\Omega_{e_{1}}}}\Gamma\left(m_{e_{1}}\right)\Gamma\left(m_{e_{1}}+\frac{1}{2}\right)\sqrt{\frac{m_{e_{2}}}{\Omega_{e_{2}}}}\Gamma\left(m_{e_{2}}\right)\Gamma\left(m_{e_{2}}+\frac{1}{2}\right)}$, 
and {$\upsilon_{2}=a_{e}+n_{2}+1$}. For the first hop of $\mathcal{T}-\mathcal{I}_{E}-\mathcal{E}$ link, $m_{e_{1}}$ and $\Omega_{e_{1}}$ denote the fading parameter and scale parameter (second moment or mean power), respectively, whereas for the second hop, they are denoted by $m_{e_{2}}$ and $\Omega_{e_{2}}$, respectively. Note that for $S_{1}=S_{2}$, $m_{r_{1}}=m_{e_{1}}$, and $\Omega_{r_{1}}=\Omega_{e_{1}}$, \textit{Scenario-I} reduces to \textit{Scenario-II}.

\subsection{CDF of SNR for Dual-hop RIS-UOWC Link}

The CDF of $\gamma_{eq}$ is expressed as \cite[Eq.~(15)]{odeyemi2019impact}
\begin{align}
\label{eq9}
\nonumber
F_{\gamma_{eq}}(\gamma)&=\text{Pr}\left\{\text{min}(\gamma_{\mathcal{R}}, \gamma_{\mathcal{U}})<\gamma\right\}
\\
&=F_{\gamma_{\mathcal{R}}}(\gamma)+F_{\gamma_{\mathcal{U}}}(\gamma)-F_{\gamma_{\mathcal{R}}}(\gamma)F_{\gamma_{\mathcal{U}}}(\gamma).
\end{align} 
Substituting \eqref{eq4} and \eqref{eq6} in \eqref{eq9} and performing mathematical manipulations, the simplified CDF of $\gamma_{eq}$ is obtained as
\begin{align}\label{eq10}
\nonumber
F_{\gamma_{eq}}(\gamma)&=\sum_{n_{1}=0}^{\infty}K_{r_{4}}\gamma^{\frac{\upsilon_{1}}{2}}+\sum_{i=1}^{2}\zeta_{i}\,G_{1,2}^{1,1}\left[{\Psi}_{i}\gamma^{{\Lambda}_{i}}\biggl|
\begin{array}{c}
1 \\
\chi_{i},0 \\
\end{array}
\right]
\\
&\times\left(1-\sum_{n_{1}=0}^{\infty}K_{r_{4}}\gamma^{\frac{\upsilon_{1}}{2}}\right).
\end{align}
As per the comprehension discussed in the literature review section, it is noticeable that the combination of RIS-aided RF-UOWC framework considering Nakagami-$m$ and mEGG distribution is not delineated in any existing research literature yet. Hence, the expression in \eqref{eq10} is attested to be a novel expression. Also, generalized characterisation of both Nakagami-$m$ and mEGG distribution leads this work to unify the existing models as special cases.

\section{Performance Analysis}

In this section, we derive the expressions of the performance measures i.e. ASC, exact and lower bound of SOP, and SPSC of the proposed RIS-aided RF-UOWC network utilizing \eqref{eq7}, \eqref{eq8}, and \eqref{eq10}. 

\subsection{Average Secrecy Capacity Analysis}

ASC is the average value of the instantaneous secrecy capacity that is mathematically interpreted as \cite[Eq.~(15)]{islam2020secrecy}
\begin{align}
\label{eq11}
\mathrm{ASC}=\int_{0}^{\infty}\frac{1}{1+\gamma}\,F_{\gamma_{\mathcal{E}}}(\gamma)\,\left[1-F_{\gamma_{eq}}(\gamma)\right]\,d\gamma.
\end{align} 
On substituting \eqref{eq8} and \eqref{eq10} into \eqref{eq11}, ASC is derived as 
\begin{align}
\label{asc1}
\nonumber
ASC&=\sum_{n_{2}=0}^{\infty}K_{e_{4}}\,\xi_{1}-\sum_{n_{1}=0}^{\infty}\sum_{n_{2}=0}^{\infty}K_{r_{4}}K_{e_{4}}\,\xi_{2}
\\
&-\sum_{n_{2}=0}^{\infty}\sum_{i=1}^{2}\zeta_{i}\,K_{e_{4}}\left(\xi_{3}-\sum_{n_{1}=0}^{\infty}K_{r_{4}}\,\xi_{4}\right),
\end{align} 
where $\xi_{1}$, $\xi_{2}$, $\xi_{3}$, and $\xi_{4}$ are the integral parts that are derived as follows.

\subsubsection{Derivation of $\xi_{1}$}

$\xi_{1}$ is expressed as
\begin{align}
\label{eq12}
\xi_{1}=\int_{0}^{\infty}\frac{\gamma^{\frac{\upsilon_{2}}{2}}}{1+\gamma}\,d\gamma.
\end{align} 
Exploiting the identity \cite[Eq.~(3.194.3)]{gradshteyn2014table}, $\xi_{1}$ is obtained as
\begin{align}
\xi_{1}=\mathcal{B}\left(\frac{\upsilon_{2}}{2}+1,-\frac{\upsilon_{2}}{2}\right),
\end{align} 

where $\mathcal{B}\left(.,.\right)$ represents the well known Beta function \cite[Eq.~(8.39)]{gradshteyn2014table}.


\subsubsection{Derivation of $\xi_{2}$}

$\xi_{2}$ is expressed as
\begin{align}
\xi_{2}=\int_{0}^{\infty}\frac{\gamma^{\frac{\upsilon_{1}+\upsilon_{2}}{2}}}{1+\gamma}d\gamma.
\end{align} 
Following the same identity utilized to derive $\xi_{1}$, $\xi_{2}$ is closed in as
\begin{align}
\xi_{2}=\mathcal{B}\left(\frac{\upsilon_{1}+\upsilon_{2}}{2}+1,-\frac{\upsilon_{1}+\upsilon_{2}}{2}\right).
\end{align}

\subsubsection{Derivation of $\xi_{3}$}

$\xi_{3}$ is expressed as
\begin{align}
\xi_{3}=\int_{0}^{\infty}\frac{\gamma^{\frac{\upsilon_{2}}{2}}}{1+\gamma}\,G_{1,2}^{1,1}\left[\Psi_{i}\gamma^{\Lambda_{i}}\biggl|
\begin{array}{c}
 1 \\
 \chi_{i},0 \\
\end{array}
\right]d\gamma.
\end{align} 
On exploiting identity \cite[Eq.~(8.4.2.5)]{prudnikov2003integrals}, the function $\frac{1}{1+\gamma}$ is transformed into Meijer's $G$ function and solving the integral with the aid of \cite[Eq.~(2.24.1.1)]{prudnikov2003integrals}, $\xi_{3}$ is obtained as
\begin{align}
\nonumber
\xi_{3}&=\int_{0}^{\infty}\gamma^{\frac{\upsilon_{2}}{2}}\,G_{1,1}^{1,1}\left[\gamma\biggl|
\begin{array}{c}
 0 \\
 0 \\
\end{array}
\right]\,G_{1,2}^{1,1}\left[\Psi_{i}\gamma^{\Lambda_{i}}\biggl|
\begin{array}{c}
 1 \\
 \chi_{i},0 \\
\end{array}
\right]d\gamma
\\
&=\frac{1}{(2\pi)^{\Lambda_{i}-1}}\,G_{1+\Lambda_{i},1+2\Lambda_{i}}^{1+\Lambda_{i},1+\Lambda_{i}}\left[\Psi_{i}\biggl|
\begin{array}{c}
 \Delta(1,1), J_{1} \\
 \Delta(1,\chi_{i}), J_{1}, \Delta(1,0) \\
\end{array}
\right],
\end{align}
where $J_{1}=\Delta\left(\Lambda_{i},-\frac{\upsilon_{2}}{2}\right)$ that follows the sequence: $\Delta(y,z)=\frac{z}{y}, \frac{z+1}{y}, ..., \frac{z+y-1}{y}$ as defined in \cite[Eq.~(22)]{adamchik1990algorithm}.
\subsubsection{Derivation of $\xi_{4}$}

Lastly, $\xi_{4}$ is setup as
\begin{align}
\xi_{4}=\int_{0}^{\infty}\frac{\gamma^{\frac{\upsilon_{1}+\upsilon_{2}}{2}}}{1+\gamma}\,G_{1,2}^{1,1}\left[\Psi_{i}\gamma^{\Lambda_{i}}\biggl|
\begin{array}{c}
 1 \\
 \chi_{i},0 \\
\end{array}
\right]d\gamma.
\end{align} 
Employing similar process as in deriving $\xi_{3}$ is carried out and $\xi_{4}$ is integrated as
\begin{align}
\xi_{4}=\frac{1}{(2\pi)^{\Lambda_{i}-1}}\,G_{1+\Lambda_{i},1+2\Lambda_{i}}^{1+\Lambda_{i},1+\Lambda_{i}}\left[\Psi_{i}\biggl|
\begin{array}{c}
 \Delta(1,1), J_{2} \\
 \Delta(1,\chi_{i}), J_{2}, \Delta(1,0) \\
\end{array}
\right],
\end{align}
where $J_{2}=\Delta\left(\Lambda_{i},-\frac{\upsilon_{1}+\upsilon_{2}}{2}\right)$.

\subsection{Secrecy Outage Probability Analysis}

A perfect secrecy is achieved only if the value of instantaneous secrecy capacity, $C_{s}$, is greater than a predecided target secrecy rate, $\epsilon_{0}$ i.e. $\epsilon_{0}\leq C_{s}$. An outage occurs when $C_{s}$ falls below $\epsilon_{0}$. The exact SOP of a mixed RIS-aided UOWC network in the appearance of an eavesdropper can be defined as \cite[Eq.~(20)]{sarker2020secrecy}
\begin{align}
\label{eqn:sope}
\nonumber
SOP_{E}&=\text{Pr}\left\{C_{s}\leq \epsilon_{o}\right\}=\text{Pr}\left\{\gamma_{eq}\leq\phi\,\gamma_{\mathcal{E}}+\phi-1\right\}
\\
&=\int_{0}^{\infty}F_{\gamma_{eq}}(\phi\,\gamma+\phi-1)\,f_{\gamma_{\mathcal{E}}}(\gamma)\,d\gamma,
\end{align}
where $\phi=2^{\epsilon_{o}}$. Substituting \eqref{eq7} and \eqref{eq10} into \eqref{eqn:sope}, exact SOP is derived as
\begin{align}
\label{eqn:sopextfinal}
SOP=\sum_{n_{1}=0}^{\infty}\sum_{q_{1}=0}^{\infty}\sum_{p_{1}=0}^{q_{1}}K_{r_{6}}\left(X_{1}-\sum_{i=1}^{2}\zeta_{i}X_{3}\right)+\sum_{i=1}^{2}\zeta_{i}K_{e_{1}}X_{2},
\end{align} 
where ${X}_{1}$, ${X}_{2}$, and $X_{3}$ are three integral parts that are derived as follows.

\subsubsection{Derivation of ${X}_{1}$}
${X}_{1}$ is expressed as
\begin{align}
\label{eqn:sope1}
{X}_{1}&=\int_{0}^{\infty}\left(\phi-1+\phi\,\gamma\right)^{\frac{\upsilon_{1}}{2}}\gamma^{K_{{e}_{2}}}\,e^{-K_{{e}_{3}}\gamma^{\frac{1}{2}}}d\gamma.
\end{align} 
Proceeding with some mathematical manipulations defined in \cite{moualeu2019physical}, and applying binomial theorem \cite[Eq.~(1.110) and Eq.~(1.111) ]{zwillinger2007table} and identity \cite[Eq.~(3.351.3)]{gradshteyn2014table}, \eqref{eqn:sope1} is integrated as
\begin{align}
\nonumber
X_{1}&=\int_{0}^{\infty}\gamma^{p_{1}+K_{{e}_{2}}}\,e^{-K_{{e}_{3}}\gamma^{\frac{1}{2}}}d\gamma
\\
\nonumber
&=2\int_{0}^{\infty}\gamma^{2\,p_{1}+2K_{{e}_{2}}+1}\,e^{-K_{{e}_{3}}\gamma}\,d\gamma
\\
&=2\,(2\,p_{1}+2K_{{e}_{2}}+1)!\,K_{{e}_{3}}^{-2(p_{1}+K_{{e}_{2}}+1)}.
\end{align}


\subsubsection{Derivation of ${X}_{2}$}

Likewise, $X_{2}$ is illustrated as
\begin{align}
\nonumber
{X}_{2}=\int_{0}^{\infty}\gamma^{K_{{e}_{2}}}e^{-K_{{e}_{3}}\gamma^{\frac{1}{2}}}G_{1,2}^{1,1}\left[{\Psi}_{i}(\phi-1+\phi\,\gamma)^{{\Lambda}_{i}}\biggl|
\begin{array}{c}
 1 \\
 {\chi}_{i},0 \\
\end{array}
\right]\,d\gamma.
\end{align} 
Utilizing the identities \cite[Eq.~(2.24.1.1)]{prudnikov2003integrals} and applying binomial theorem \cite[Eq.~(1.111)]{zwillinger2007table}, ${X}_{2}$ is expressed in an alternative form and derived as
\begin{align}
&X_{2}=\int_{0}^{\infty}\gamma^{K_{{e}_{2}}}e^{-K_{{e}_{3}}\gamma^{\frac{1}{2}}}G_{1,2}^{1,1}\left[F_{2}\,\gamma^{p_{2}}\biggl|
\begin{array}{c}
 1 \\
 {\chi}_{i},0 \\
\end{array}
\right]\,d\gamma
\\
\nonumber
&=2\int_{0}^{\infty}\gamma^{2K_{{e}_{2}}+1}\,G_{0,1}^{1,0}\left[K_{{e}_{3}}\,\gamma\biggl|
\begin{array}{c}
 - \\
 0 \\
\end{array}
\right]G_{1,2}^{1,1}\left[F_{2}\,\gamma^{2\,p_{2}}\biggl|
\begin{array}{c}
 1 \\
 {\chi}_{i},0 \\
\end{array}
\right]d\gamma
\\
&=\frac{(2\,p_{2})^{2K_{{e}_{2}}+\frac{3}{2}}K_{{e}_{3}}^{-2K_{{e}_{2}}-2}}{(2\pi)^{p_{2}-\frac{1}{2}}}\,G_{1+2\,p_{2},2}^{1,1+2\,p_{2}}\biggl[\frac{F_{2}\,(2\,p_{2})^{2\,p_{2}}}{K_{e_{3}}^{2\,p_{2}}}\biggl|
\begin{array}{c}
1, J_{3}\\
\chi_{i},0\\
\end{array}
\biggl],
\end{align}


where $F_{2}={\Psi}_{i}\sum_{p_{2}=0}^{{\Lambda}_{i}}\binom{\Lambda_{i}}{p_{2}}(\phi-1)^{\Lambda_{i}-p_{2}}\,\phi^{\,p_{2}}$ and $J_{3}=\Delta\left(2\,p_{2},-1-2K_{{e}_{2}}\right)$.
\subsubsection{Derivation of ${X}_{3}$}

Now, $X_{3}$ is illustrated as
\begin{align}
\nonumber
{X}_{3}&=\int_{0}^{\infty}\gamma^{p_{1}+K_{{e}_{2}}}e^{-K_{{e}_{3}}\gamma^{\frac{1}{2}}}
\\
&\times G_{1,2}^{1,1}\left[{\Psi}_{i}(\phi-1+\phi\gamma)^{{\Lambda}_{i}}\biggl|
\begin{array}{c}
 1 \\
 {\chi}_{i},0 \\
\end{array}
\right]d\gamma.
\end{align} 
Utilizing similar identities as utilized for ${X}_{2}$, ${X}_{3}$ is expressed in an alternative notion and integrated as
\begin{align}
\nonumber
X_{3}&=\int_{0}^{\infty}\gamma^{p_{1}+K_{{e}_{2}}}e^{-K_{{e}_{3}}\gamma^{\frac{1}{2}}}G_{1,2}^{1,1}\left[F_{2}\,\gamma^{p_{2}}\biggl|
\begin{array}{c}
 1 \\
 {\chi}_{i},0 \\
\end{array}
\right]d\gamma
\\
\nonumber
\\
\nonumber
&=2\int_{0}^{\infty}\gamma^{2\,p_{1}+2K_{{e}_{2}}+1}\,G_{0,1}^{1,0}\left[K_{{e}_{3}}\,\gamma\biggl|
\begin{array}{c}
 - \\
 0 \\
\end{array}
\right]
\\
\nonumber
&\times G_{1,2}^{1,1}\biggl[F_{2}\,\gamma^{2\,p_{2}}\biggl|
\begin{array}{c}
 1 \\
 {\chi}_{i},0 \\
\end{array}
\biggl]\,d\gamma=\frac{2\,(2\,p_{2})^{2\,p_{1}+2K_{{e}_{2}}+\frac{3}{2}}}{K_{{e}_{3}}^{2\,p_{1}+2K_{{e}_{2}}+2}(2\pi)^{p_{2}-\frac{1}{2}}}
\\
&\times G_{1+2\,p_{2},2}^{1,1+2\,p_{2}}\left[\frac{F_{2}\,(2\,p_{2})^{2\,p_{2}}}{K_{e_{3}}^{2\,p_{2}}}\biggl|
\begin{array}{c}
 1, J_{4} \\
\chi_{i}, 0 \\
\end{array}
\right],
\end{align} 


where $J_{4}=\Delta(2\,p_{2}, -1-2\,p_{1}-2K_{{e}_{2}})$.

\subsection*{\textbf{Lower Bound of Secrecy Outage Probability Analysis:}}

As per \cite{badrudduza2021security}, the lower bound of SOP can be setup as
\begin{small}
\begin{align}
\label{eqn:sop1}
SOP\geq SOP_{L}=\text{Pr}\left\{\gamma_{eq}\leq\phi\,\gamma_{\mathcal{E}}\right\}=\int_{0}^{\infty}F_{\gamma_{eq}}(\phi\,\gamma)\,f_{\gamma_{\mathcal{E}}}(\gamma)\,d\gamma.
\end{align}
\end{small}
By substituting \eqref{eq7} and \eqref{eq10} into \eqref{eqn:sop1}, lower bound SOP is derived as
\begin{align}
\label{eqn:sop3}
SOP_{L}=\sum_{n_{1}=0}^{\infty}K_{r_{5}}R_{1}+\sum_{i=1}^{2}\zeta_{i}\left(K_{e_{1}}R_{2}-\sum_{n_{1}=0}^{\infty}K_{r_{5}}R_{3}\right),
\end{align}
where $K_{r_{5}}=K_{r_{4}}K_{e_{1}}\phi^{\frac{\upsilon_{1}}{2}}$ and derivations of the three integral terms $R_{1}$, $R_{2}$, and $R_{3}$ are expressed as follows.

\subsubsection{Derivation of $R_{1}$}

$R_{1}$ is expressed as
\begin{align}
R_{1}=\int_{0}^{\infty}\gamma^{\frac{\upsilon_{1}}{2}+K_{{e}_{2}}}\,e^{-K_{{e}_{3}}\gamma^{\frac{1}{2}}}\,d\gamma.
\end{align} 
Employing identities \cite[Eqs.~(3.381.10) and (8.356.3)]{gradshteyn2014table}, $R_{1}$ is derived as
\begin{align}
R_{1}=2\,\Gamma\left(\upsilon_{1}+2K_{e_{2}}+2\right)/K_{e_{3}}^{\upsilon_{1}+2K_{e_{2}}+2}.
\end{align}

\subsubsection{Derivation of $R_{2}$}

$R_{2}$ is expressed as
\begin{align}
R_{2}=\int_{0}^{\infty}\gamma^{K_{{e}_{2}}}\,e^{-K_{{e}_{3}}\gamma^{\frac{1}{2}}}G_{1,2}^{1,1}\left[{\Psi}_{i}(\phi\,\gamma)^{\Lambda_{i}}\biggl|
\begin{array}{c}
1 \\
\chi_{i},0 \\
\end{array}
\right]d\gamma,
\end{align}

where $\phi=2^{\epsilon_{0}}$  Now, $R_{2}$, with the help of some mathematical manipulations in \cite[Eq.~(8.4.3.1) and (2.24.1.1)]{prudnikov2003integrals}, is derived as
\begin{small}
\begin{align}
\nonumber
&R_{2}=\int_{0}^{\infty}\gamma^{2\,(K_{{e}_{2}}+1)}e^{-K_{{e}_{3}}\gamma^{\frac{1}{2}}}G_{1,2}^{1,1}\left[Z_{1}\gamma^{2\Lambda_{i}}\biggl|
\begin{array}{c}
1 \\
\chi_{i},0 \\
\end{array}
\right]d\gamma^{2}
\\
\nonumber 
&=\int_{0}^{\infty}\gamma^{2\,(K_{{e}_{2}}+1)}\,G_{0,1}^{1,0}\left[K_{{e}_{3}}\gamma\biggl|
\begin{array}{c}
 - \\
 0 \\
\end{array}
\right]G_{1,2}^{1,1}\left[Z_{1}\gamma^{2\Lambda_{i}}\biggl |
\begin{array}{c}
 1 \\
\chi_{i},0 \\
\end{array}
\right]d\gamma^{2}
\\
&=Z_{2}\,G_{1+2\Lambda_{i},2}^{1,1+2\Lambda_{i}}\left[Z_{1}Z_{3}\biggl|
\begin{array}{c}
\Delta(1,1), J_{5} \\
\Delta(1,\chi_{i}), \Delta(1,0) \\
\end{array}\right],
\end{align}
\end{small}
where $Z_{1}={\Psi}_{i}\phi^{\Lambda_{i}}, Z_{2}=\frac{2\,(2\Lambda_{i})^{2K_{e_{2}}+\frac{3}{2}}}{K_{e_{3}}^{2K_{e_{2}}+2}\,(2\pi)^{\Lambda_{i}-\frac{1}{2}}}$, $Z_{3}=\left(\frac{2\Lambda_{i}}{K_{{e}_{3}}}\right)^{2\Lambda_{i}}$, and $J_{5}=\Delta(2\Lambda_{i}, -1-2K_{{e}_{2}})$.

\subsubsection{Derivation of $R_{3}$}

$R_{3}$ is expressed as
\begin{align}
\nonumber
R_{3}&=\int_{0}^{\infty}\gamma^{\frac{\upsilon_{1}}{2}+K_{{e}_{2}}}\,e^{-K_{{e}_{3}}\gamma^{\frac{1}{2}}}G_{1,2}^{1,1}\left[{\Psi}_{i}(\phi\,\gamma)^{\Lambda_{i}}\biggl|
\begin{array}{c}
1 \\
\chi_{i},0 \\
\end{array}\right]d\gamma.
\end{align} 
With the help of similar process followed while deriving $R_{2}$, $R_{3}$ is finally derived as
\begin{align}
R_{3}=Z_{2}\left(\frac{2\Lambda_{i}}{K_{e_{3}}}\right)^{\upsilon_{1}}G_{1+2\Lambda_{i},2}^{1,1+2\Lambda_{i}}\left[Z_{1}Z_{3}\biggl|
\begin{array}{c}
\Delta(1,1), J_{6} \\
\Delta(1,\chi_{i}), \Delta(1,0) \\
\end{array}
\right].
\end{align} 
where $J_{6}=\Delta\left(2\Lambda_{i},-1-2K_{{e}_{2}}-\frac{\upsilon_{1}}{2}\right)$.

\subsection{Strictly Positive Secrecy Capacity Analysis}

To ensure seamless communication in a wiretapped paradigm, SPSC is a widely used performance measure that is achieved only if the secrecy capacity holds a positive quantity. According to \cite[Eq.~(25)]{islam2020secrecy}, SPSC is defined as
\begin{small}
\begin{align}
\label{eqn:spsc4}
SPSC&=\text{Pr}\left\{C_{s}>0\right\}=1-\text{Pr}\left\{C_{s}\leq0\right\}=1-SOP\left(\epsilon_{o}=0\right).
\end{align}
\end{small}
On substituting $\epsilon_{0}=0$ in \eqref{eqn:sopextfinal}, the expression of SPSC in analytical form is obtained as expressed in \eqref{eqn:spscfinal}.
\begin{figure*}[!t]
\begin{align}
\label{eqn:spscfinal}
\nonumber
SPSC&=1-\frac{\gamma_{m_{2}}^{-\frac{a_{e}+1}{2}}}{(2\,b_{e})^{a_{e}+1}\,\Gamma(a_{e}+1)}\sum_{n_{1}=0}^{\infty}\frac{(-1)^{n_{1}}K_{{r}_{3}}^{\upsilon_{1}}}{n_{1}!\,\upsilon_{1}}\frac{2\,\Gamma(\upsilon_{1}+2K_{e_{2}}+2)}{K_{e_{3}}^{\upsilon_{1}+2K_{e_{2}}+2}}-\sum_{i=1}^{2}{\zeta_{i}}\,\frac{\gamma_{m_{2}}^{-\frac{a_{e}+1}{2}}}{(2\,b_{e})^{a_{e}+1}\,\Gamma(a_{e}+1)}
\\
\nonumber
&\times\left[\frac{2\,(2\Lambda_{i})^{2K_{e_{2}}+\frac{3}{2}}}{K_{e_{3}}^{2K_{e_{2}}+2}(2\pi)^{\Lambda_{i}-\frac{1}{2}}}\,G_{1+2\Lambda_{i},2}^{1,1+2\Lambda_{i}}\left[\Psi_{i}\left(\frac{2\Lambda_{i}}{K_{{e}_{3}}}\right)^{2\Lambda_{i}}\biggl|
\begin{array}{c}
\Delta(1,1), J_{5} \\
\Delta(1,\chi_{i}), \Delta(1,0) \\
\end{array}\right]
-\sum_{n_{1}=0}^{\infty}\frac{(-1)^{n_{1}}K_{{r}_{3}}^{\upsilon_{1}}}{n_{1}!\,\upsilon_{1}}\right.
\\
&\left.\times\frac{2\,(2\Lambda_{i})^{\upsilon_{1}+2K_{e_{2}}+\frac{3}{2}}}{K_{e_{3}}^{\upsilon_{1}+2K_{e_{2}}+2}(2\pi)^{\Lambda_{i}-\frac{1}{2}}}\,G_{1+2\Lambda_{i},2}^{1,1+2\Lambda_{i}}\left[\Psi_{i} \left(\frac{2\Lambda_{i}}{K_{{e}_{3}}}\right)^{2\Lambda_{i}}\biggl|
\begin{array}{c}
\Delta(1,1), J_{6} \\
\Delta(1,\chi_{i}),\Delta(1,0) \\
\end{array}\right]
\right]
\end{align}
\hrulefill
\end{figure*}

\subsection{Generality of ASC, Exact and Lower bound of SOP, and SPSC Expressions}

With the aim of ensuring a secured RIS-aided UOWC system, we derive the expressions of ASC, exact SOP, lower bound of SOP, and SPSC as performance measures. Based on aforementioned research literature and authors' knowledge, our derived expressions in \eqref{asc1}, \eqref{eqn:sopextfinal}, \eqref{eqn:sop3}, and \eqref{eqn:spscfinal}, respectively, are attested as novel. This work is the first research work in the literature that investigates a dual-hop RIS-aided UOWC network in the presence of eavesdropping surveillance. For a special case with $r=1$, the PDF of mEGG distribution in \eqref{eq5} leads to the PDF of exponential Gamma (EG) model \cite[Eq.~(14)]{illi2019physical}. In addition, the PDF of Nakagami-$m$ distribution can be transformed into the PDFs of Rayleigh and Gaussian distributions for the cases when $m_{r_{1}}=m_{r_{2}}=m_{e_{1}}=m_{e_{2}}=1$ and $m_{r_{1}}=m_{r_{2}}=m_{e_{1}}=m_{e_{2}}=0.5$, respectively, as mentioned in \cite{wongtrairat2009performance}.


\section{Numerical Results}
This section represents insights for the effects of system parameters (i.e. fading, number of reflecting elements, scale parameter, thermal gradients, detection techniques, underwater turbulence, etc.) on the secrecy performance of the offered RIS-aided UOWC system via presenting some numerical examples with figures utilizing the derived expressions in \eqref{asc1}, \eqref{eqn:sopextfinal}, \eqref{eqn:sop3}, and \eqref{eqn:spscfinal}. To validate the novel expressions, Monte-Carlo simulations are performed by initiating $10^{6}$ random samples in MATLAB that is marked by "Sim" in each figure. To generate GG distribution randomly, MATLAB function gamrnd(.) is utilized \cite{lei2017secrecy}. To illustrate the impact of UWT, values of mEGG distribution parameters corresponding to temperature gradient and thermal uniform UOWC link presented in Tables $2$ and $3$ are also utilized \cite{badrudduza2021security}.

\begin{figure}  [!ht]
\vspace{0.00mm}
    \centerline{\includegraphics[width=0.45\textwidth]{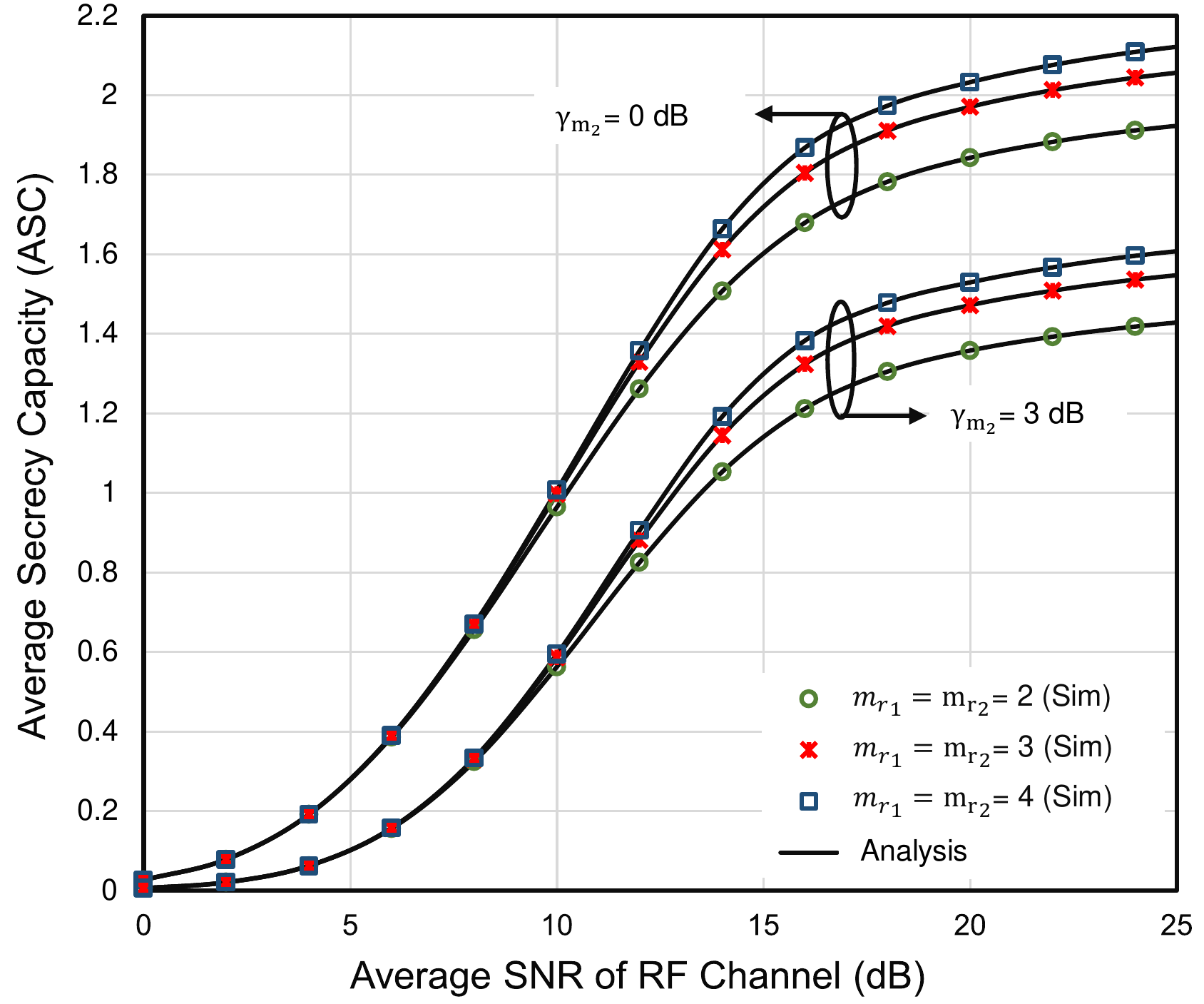}}
        \vspace{0mm}
    \caption{{\color{black}
         The ASC versus $\gamma_{m_{1}}$ for specific values of $m_{r_{1}}$, $m_{r_{2}}$, and $\gamma_{m_{2}}$ with $m_{e_{1}}=m_{e_{2}}=S_{1}=S_{2}=2$, $\Omega_{r_{1}}=\Omega_{r_{2}}=\Omega_{e_{1}}=\Omega_{e_{1}}=1$, $r=1$, $\gamma_{m_{u}}=10$ dB, $h=2.4$, and $l=0.05$.}
    }
    \label{Asc_mr1_mr2}
\end{figure}

\begin{figure}  [!ht]
\vspace{0.00mm}
    \centerline{\includegraphics[width=0.45\textwidth,angle =0]{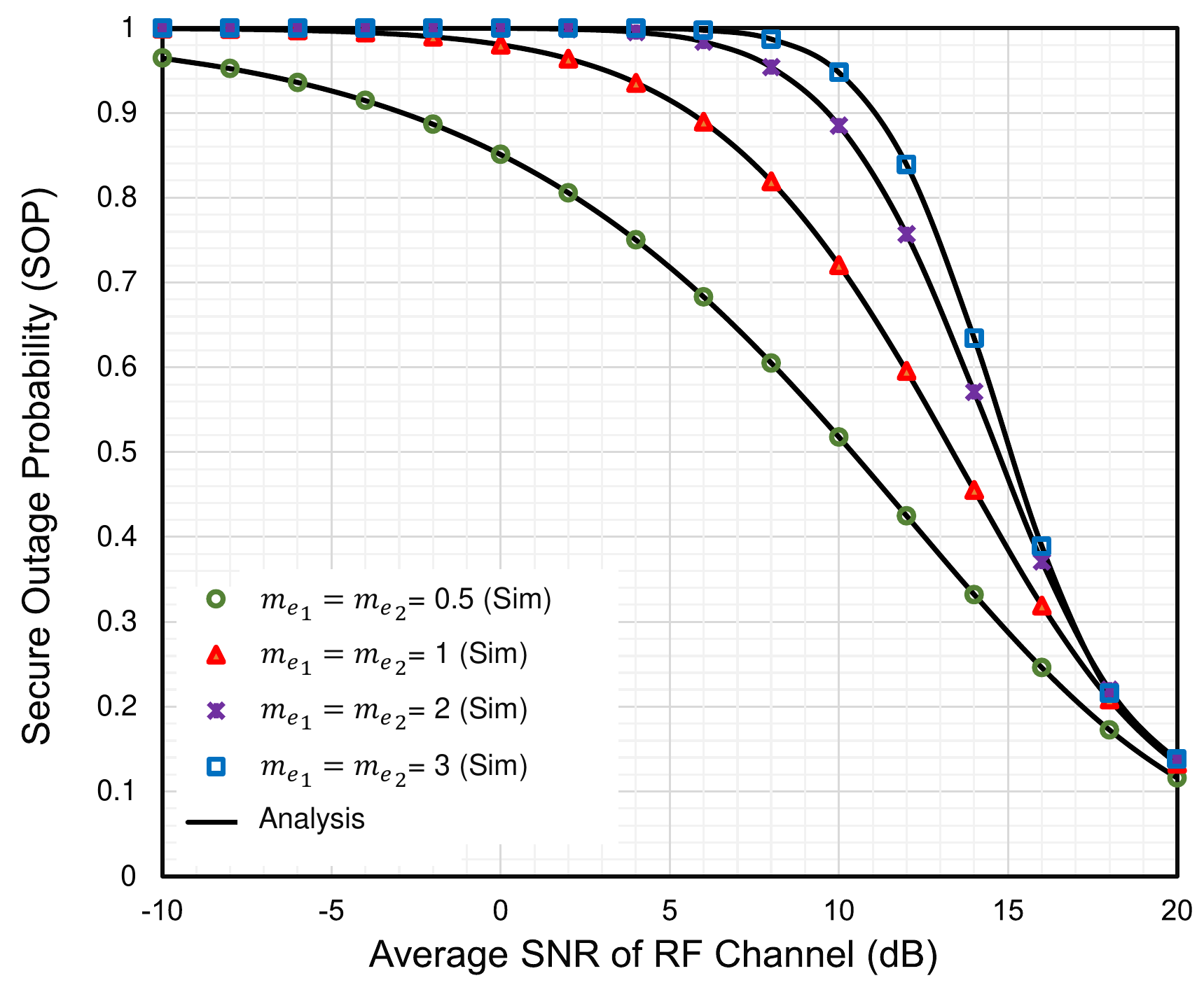}}
        \vspace{0mm}
    \caption{
         The lower bound of SOP versus $\gamma_{m_{1}}$ for specific values of $m_{e_{1}}$ and $m_{e_{2}}$ with $m_{r_{1}}=m_{r_{2}}=S_{1}=S_{2}=2$, $\Omega_{r_{1}}=\Omega_{r_{2}}=\Omega_{e_{1}}=\Omega_{e_{1}}=1$, $r=1$, $\gamma_{m_{2}}=0$ dB, $\gamma_{m_{u}}=15$ dB, $\epsilon_{0}=0.01$ bits/Hz, $h=2.4$, and $l=0.05$.
         }
    \label{SOPL_me1_me2}
\end{figure}

\begin{figure}  [!ht]
\vspace{0.00mm}
    \centerline{\includegraphics[width=0.45\textwidth,angle =0]{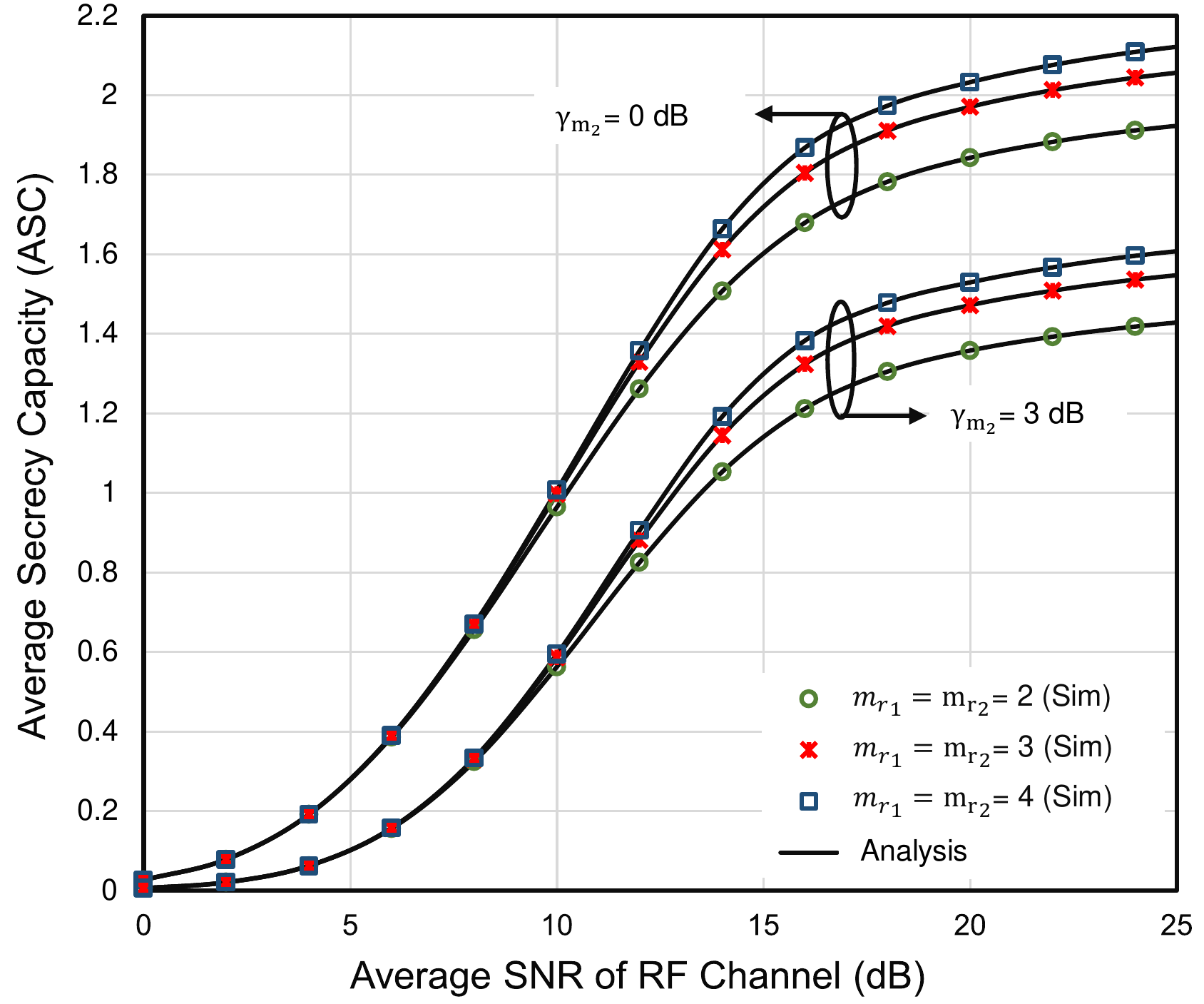}}
        \vspace{0mm}
    \caption{
         The ASC versus $\gamma_{m_{1}}$ for specific values of $m_{r_{2}}$ and $m_{e_{2}}$ with $m_{r_{1}}=m_{e_{1}}=S_{1}=S_{2}=2$, $\Omega_{r_{1}}=\Omega_{r_{2}}=\Omega_{e_{1}}=\Omega_{e_{1}}=1$, $\gamma_{m_{2}}=0$ dB, $r=1$, $\gamma_{m_{u}}=10$ dB, $h=2.4$, and $l=0.05$.
         }
\label{ASC_mr2=me2}
\end{figure}

The ASC and lower bound of SOP are plotted against $\gamma_{m_{1}}$ in Figs. \ref{Asc_mr1_mr2} and \ref{SOPL_me1_me2} considering \textit{Scenario-I} to observe and demonstrate the impact of fading severity on both the RF and eavesdropper links (assuming $m_{r_{1}}=m_{r_{2}}$ and $m_{e_{1}}=m_{e_{2}}$) under weak turbulence condition.
It is obvious from the figures that increased values of the fading parameters of $\mathcal{T}-\mathcal{I}_{R}-\mathcal{R}$ link results total fading of the corresponding link to become weaker that in turn gives rise to ASC. In contrast, increase in values of the fading parameters of $\mathcal{T}-\mathcal{I}_{E}-\mathcal{E}$ link results in deteriorated system security that is represented by poor outage performance in Fig. \ref{SOPL_me1_me2}. This result is also expected since increase in $m_{e_{1}}$ and $m_{e_{2}}$ makes $\mathcal{T}-\mathcal{I}_{E}-\mathcal{E}$ link better compared to the $\mathcal{T}-\mathcal{I}_{R}-\mathcal{R}$ link. To demonstrate the impact of fading in \textit{Scenario-II}, ASC is plotted against $\gamma_{m_{1}}$ in Fig. \ref{ASC_mr2=me2} and it is observed as expected that increase in $m_{r_{2}}$ and decrease in $m_{e_{2}}$ is beneficial for the secrecy capacity.

\begin{figure}  [!ht]
\vspace{0.00mm}
    \centerline{\includegraphics[width=0.45\textwidth,angle=0]{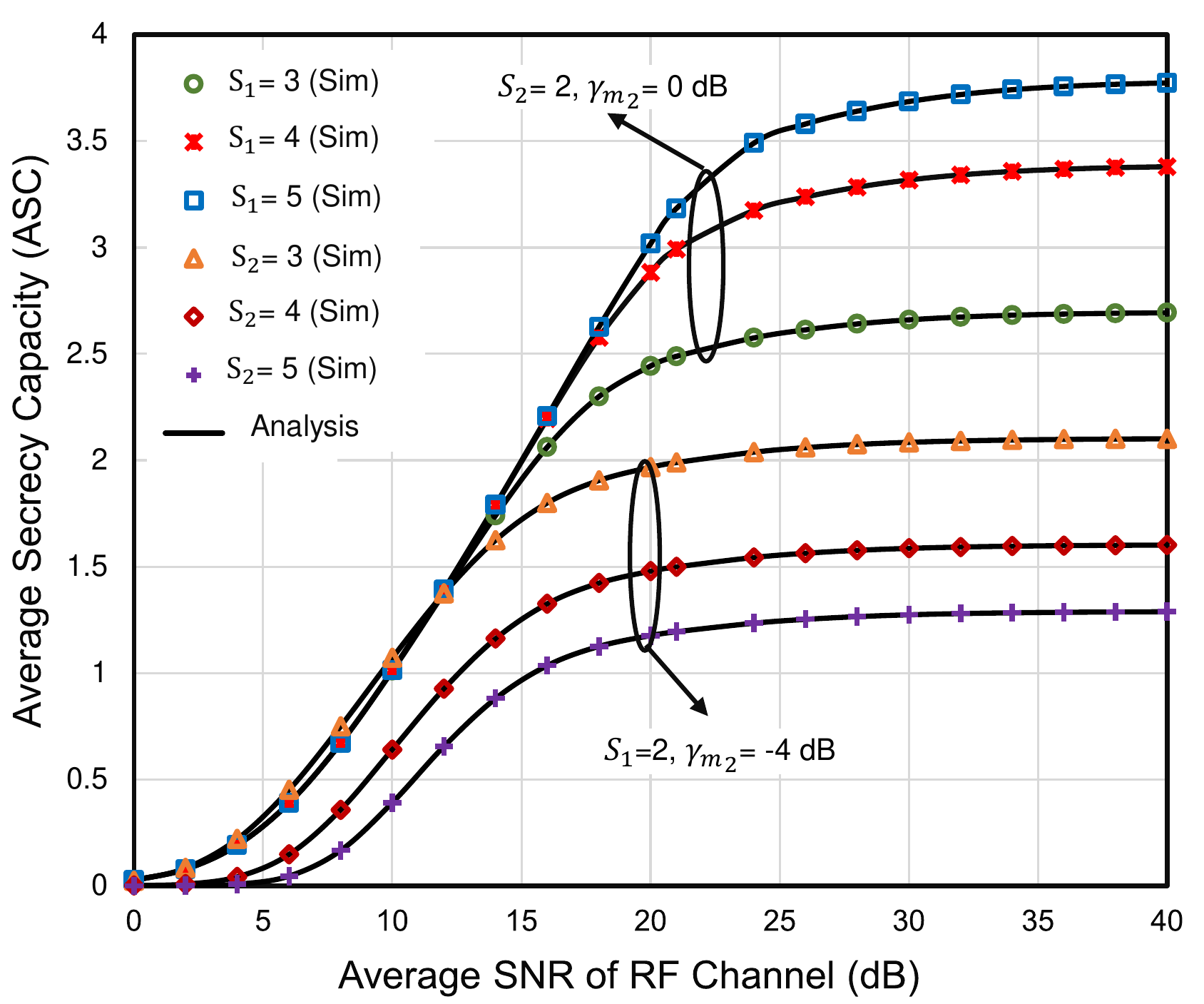}}
        \vspace{0mm}
    \caption{
         The ASC versus $\gamma_{m_{1}}$ for specific values of $S_{1}$, $S_{2}$, and $\gamma_{m_{2}}$ with $m_{r_{1}}=m_{r_{2}}=m_{e_{1}}=m_{e_{2}}=2$, $\Omega_{r_{1}}=\Omega_{r_{2}}=\Omega_{e_{1}}=\Omega_{e_{1}}=1$, $r=1$, $\gamma_{m_{u}}=10$ dB, $h=2.4$, and $l=0.05$.}
    \label{ASC_S1_S2}
\end{figure}
\begin{figure}  [!ht]
\vspace{0.00mm}
    \centerline{\includegraphics[width=0.45\textwidth,angle=0]{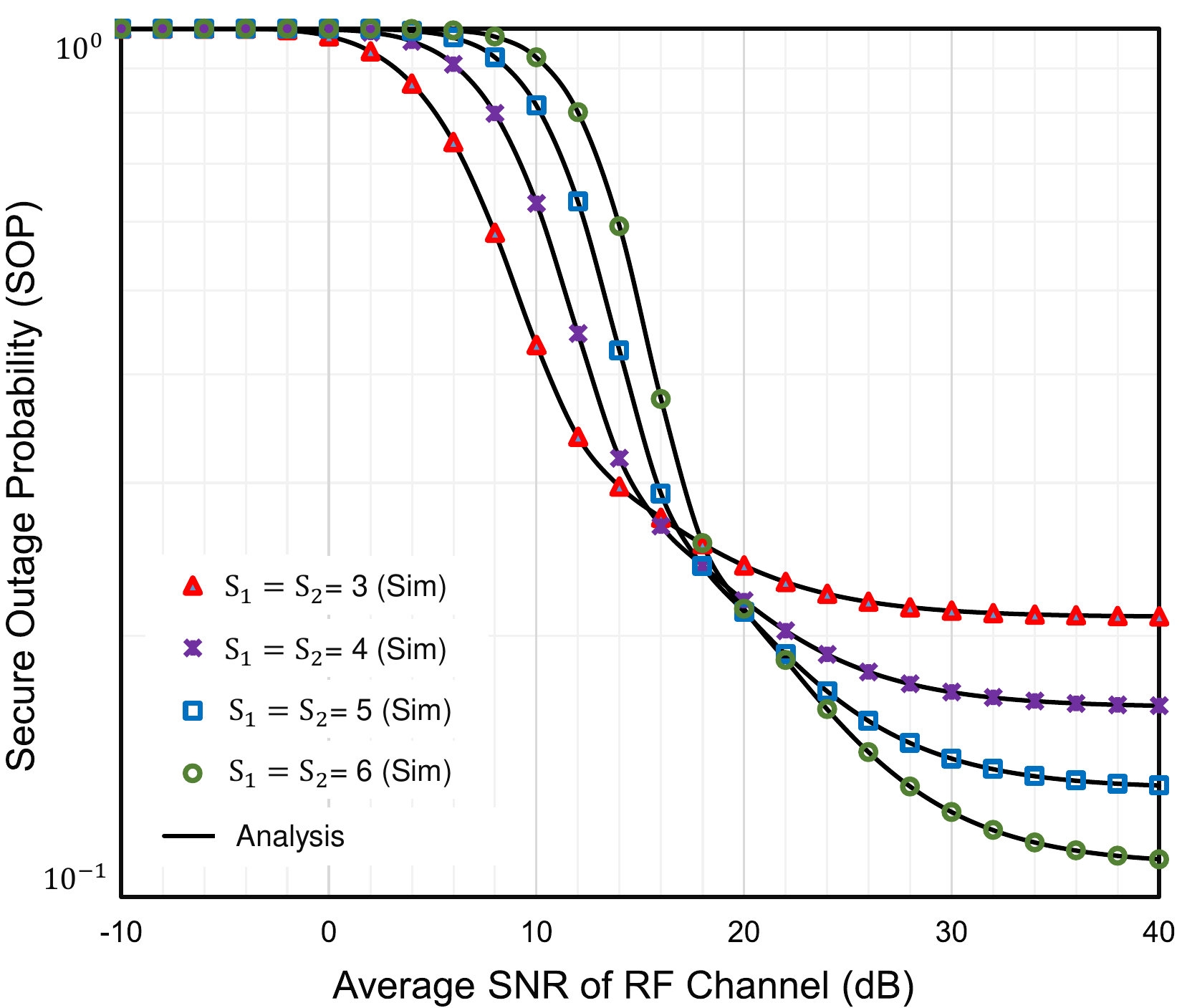}}
        \vspace{0mm}
    \caption{
         The exact SOP versus $\gamma_{m_{1}}$ for specific values of $S_{1}$ and $S_{2}$ with $m_{r_{1}}=\Omega_{r_{2}}=2$, $m_{r_{2}}=\Omega_{r_{1}}=4$, $m_{e_{1}}=m_{e_{2}}=2$, $\Omega_{e_{1}}=\Omega_{e_{1}}=1$, $r=1$, $\gamma_{m_{2}}=3$ dB, $\gamma_{m_{u}}=-5$ dB, $\epsilon_{0}=0.5$ bits/Hz, $h=2.4$, and $l=0.05$.}
    \label{SOPE_S1=S2}
\end{figure} 
\begin{figure}  [!ht]
\vspace{0.00mm}
    \centerline{\includegraphics[width=0.45\textwidth,angle =0]{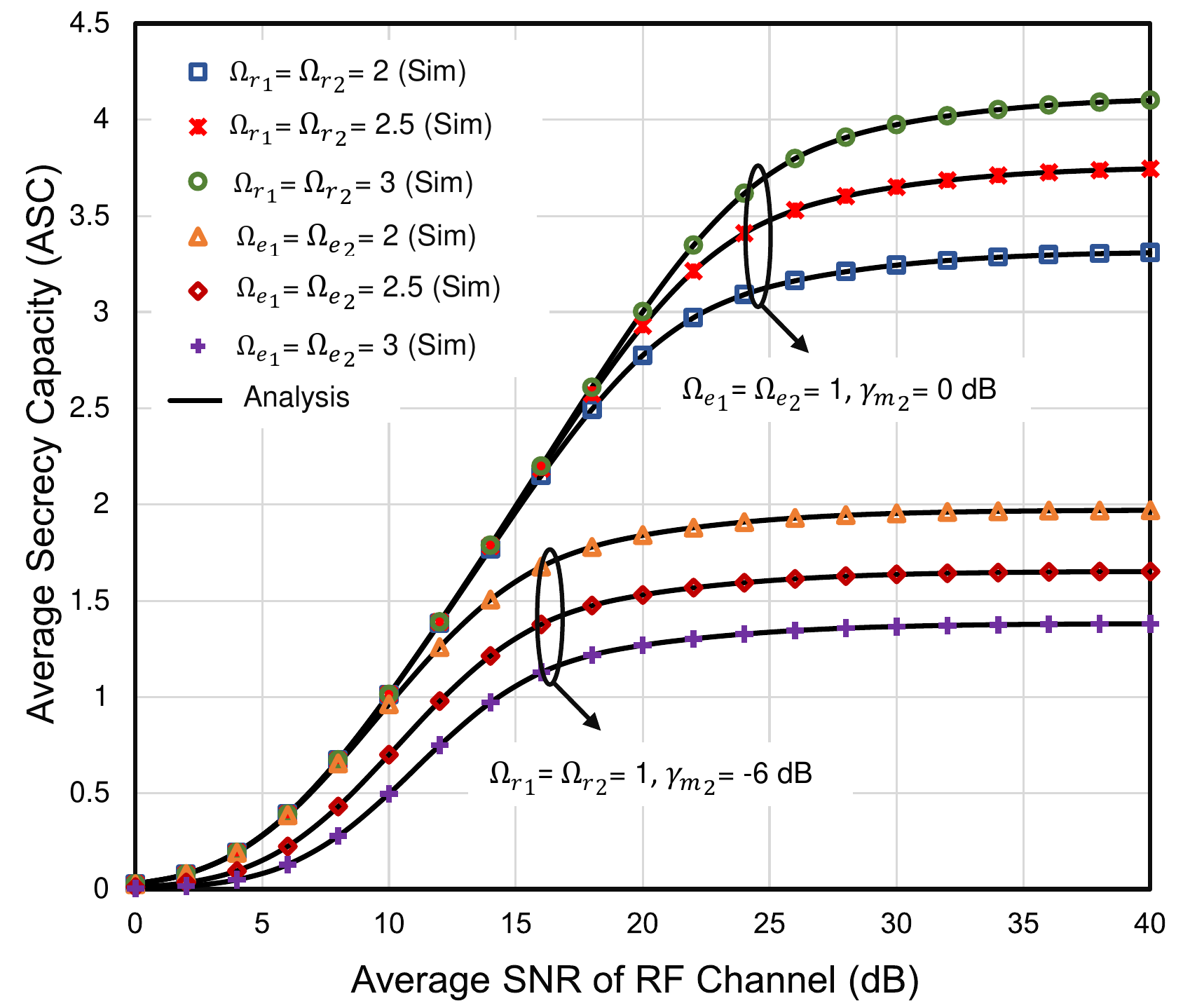}}
        \vspace{0mm}
    \caption{
         The ASC versus $\gamma_{m_{1}}$ for specific values of $\Omega_{r_{1}}$, $\Omega_{r_{2}}$, $\Omega_{e_{1}}$,  $\Omega_{e_{2}}$, and $\gamma_{m_{2}}$ with $m_{r_{1}}=m_{r_{2}}=m_{e_{1}}=m_{e_{2}}=S_{1}=S_{2}=2$,
         $r=1$, $\gamma_{m_{u}}=10$ dB, $h=2.4$, and $l=0.05$.
         }
    \label{ASC_0mega1_omega2}
\end{figure} 


Fig. \ref{ASC_S1_S2} depicts ASC versus $\gamma_{m_{1}}$ considering HD technique in \textit{Scenario-I}.
It is noted, as testified in Fig. \ref{ASC_S1_S2}, that in the presence of RIS, system performance improves with increase in $S_{1}$ and decays with $S_{2}$ as the impact of fading severity of the respective links is negligible when $S_{1}$ and $S_{2}$ are very large. Similar impacts of the reflecting elements of a RIS is also observed in Fig. \ref{SOPE_S1=S2} assuming \textit{Scenario-II} wherein exact SOP is demonstrated as a function of $\gamma_{m_{1}}$.
A closer look on Fig. \ref{SOPE_S1=S2} apprises that for a lower value of $\gamma_{m_{1}}$ ($-10$ dB to $15$ dB), system performance degrades with increasing value of $S_{1}=S_{2}$ whereas after crossing a certain value of $\gamma_{m_{1}}$ ($\sim${$18$ dB}), system regains it's desired mannerism and makes the model intractable for an eavesdropper to hack information from it. System security enhancement is also not possible as long as minimum number of reflecting elements of RIS for both the user and eavesdropper is ensured.

Besides the scale parameters, shape parameters also contributes equally in secrecy analysis that is demonstrated in Fig. \ref{ASC_0mega1_omega2} via plotting ASC versus $\gamma_{m_{1}}$ with \textit{Scenario-I}.
\begin{figure}  [!ht]
\vspace{0.00mm}
    \centerline{\includegraphics[width=0.45\textwidth,angle =0]{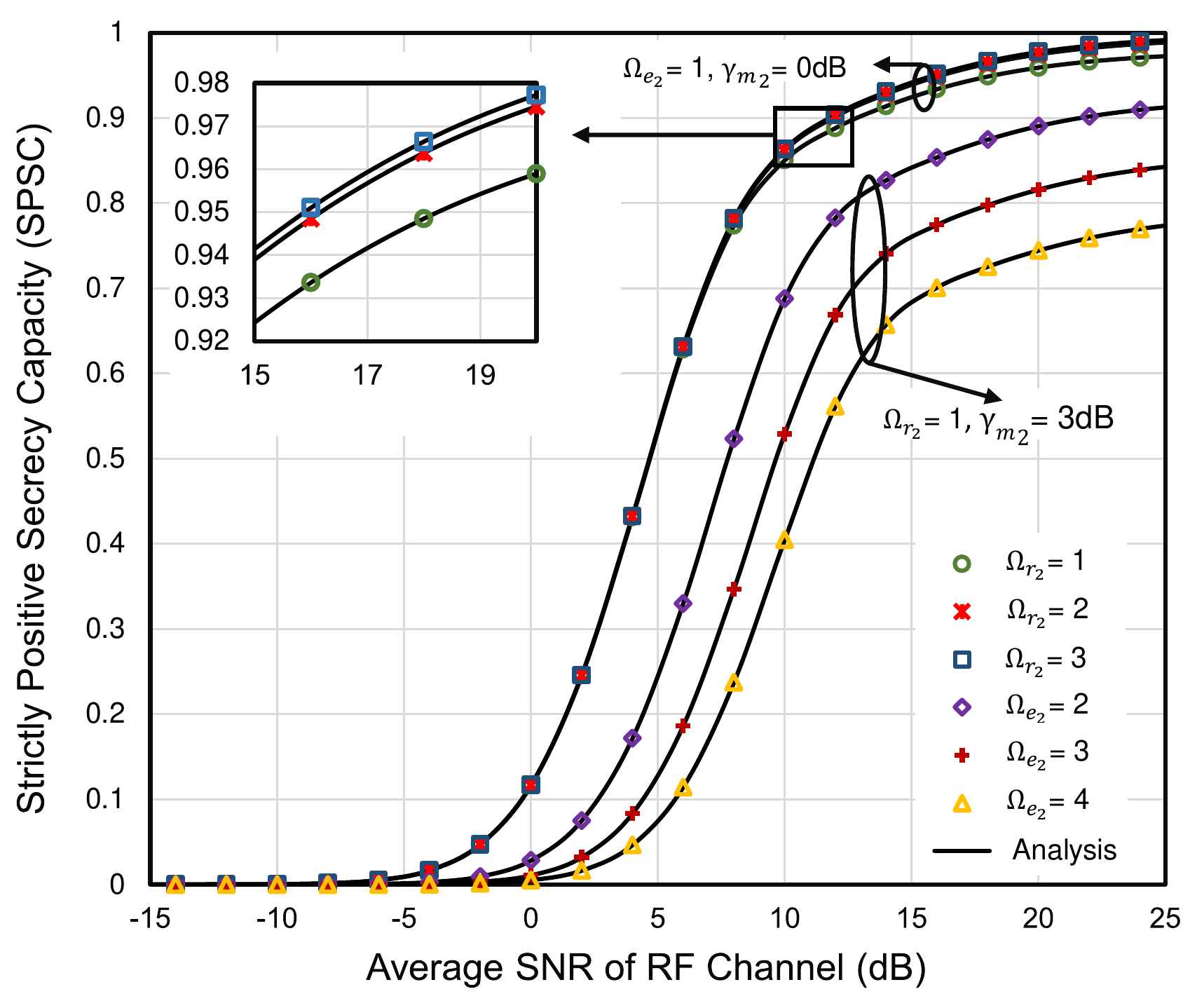}}
        \vspace{0mm}
    \caption{
         The SPSC versus $\gamma_{m_{1}}$ for specific values of $\Omega_{r_{2}}$, $\Omega_{e_{2}}$, and $\gamma_{m_{2}}$ with $m_{r_{1}}=m_{r_{2}}=m_{e_{1}}=m_{e_{2}}=S_{1}=S_{2}=2$, $\Omega_{r_{1}}=\Omega_{e_{1}}=1$, $r=1$, $\gamma_{m_{u}}=10$ dB, $h=2.4$, and $l=0.05$.}
    \label{SPSC_omgr2=omge2}
\end{figure} 
\begin{figure}  [!ht]
\vspace{0.00mm}
    \centerline{\includegraphics[width=0.45\textwidth,angle =0]{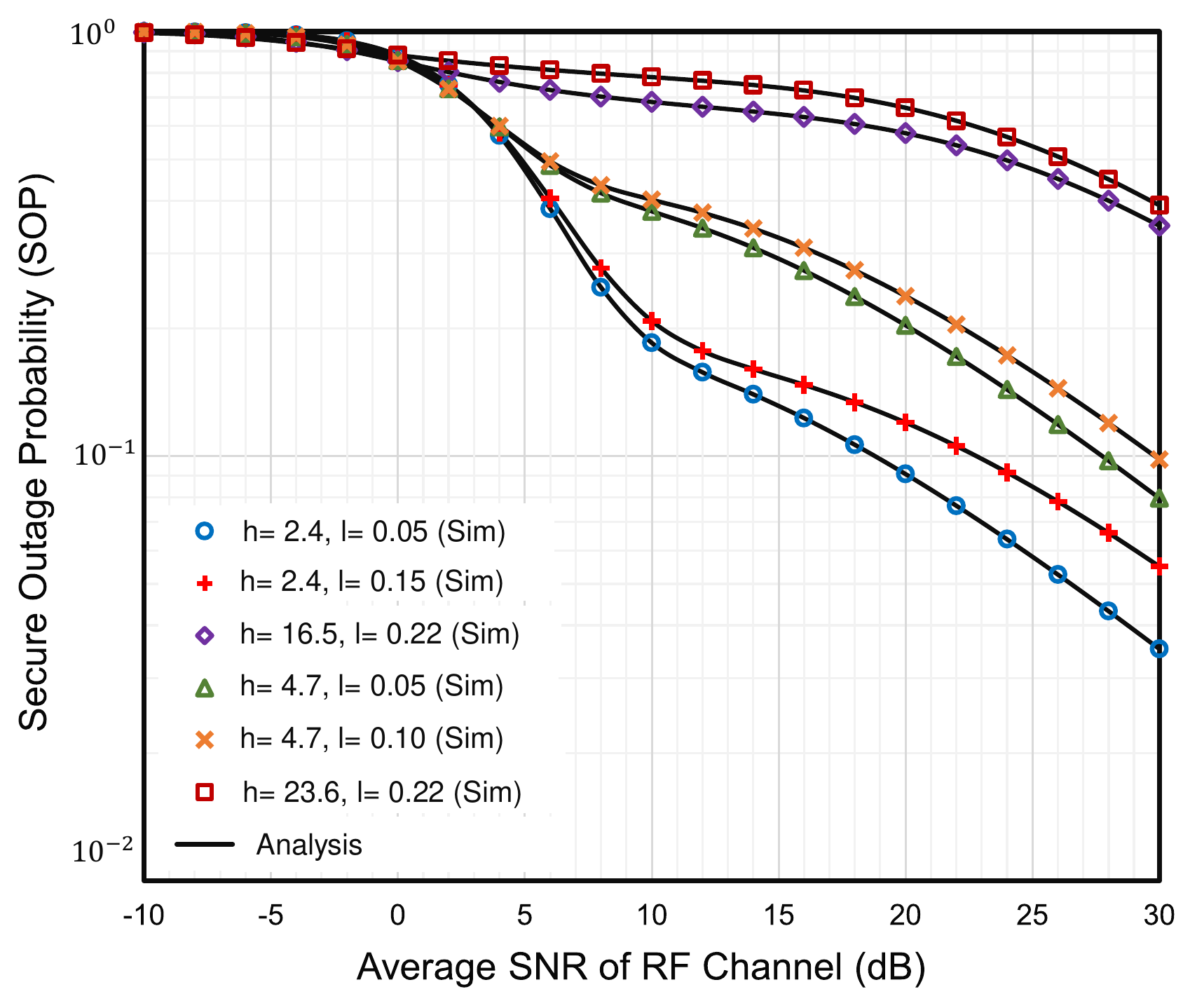}}
        \vspace{0mm}
    \caption{
          The lower bound of SOP versus $\gamma_{m_{1}}$ for specific values of $h$ and $l$, $m_{r_{1}}=m_{r_{2}}=m_{e_{1}}=m_{e_{2}}=S_{1}=S_{2}=2$, $\Omega_{r_{1}}=\Omega_{r_{2}}=\Omega_{e_{1}}=\Omega_{e_{2}}=1$, $\gamma_{m_{2}}=0$ dB,$r=1$, $\gamma_{m_{u}}=10$ dB, and $\epsilon_{0}=0.01$ bits/Hz.}
    \label{SOPL_therm_grad}
\end{figure} 
\begin{figure}  [!ht]
\vspace{0.00mm}
    \centerline{\includegraphics[width=0.45\textwidth,angle =0]{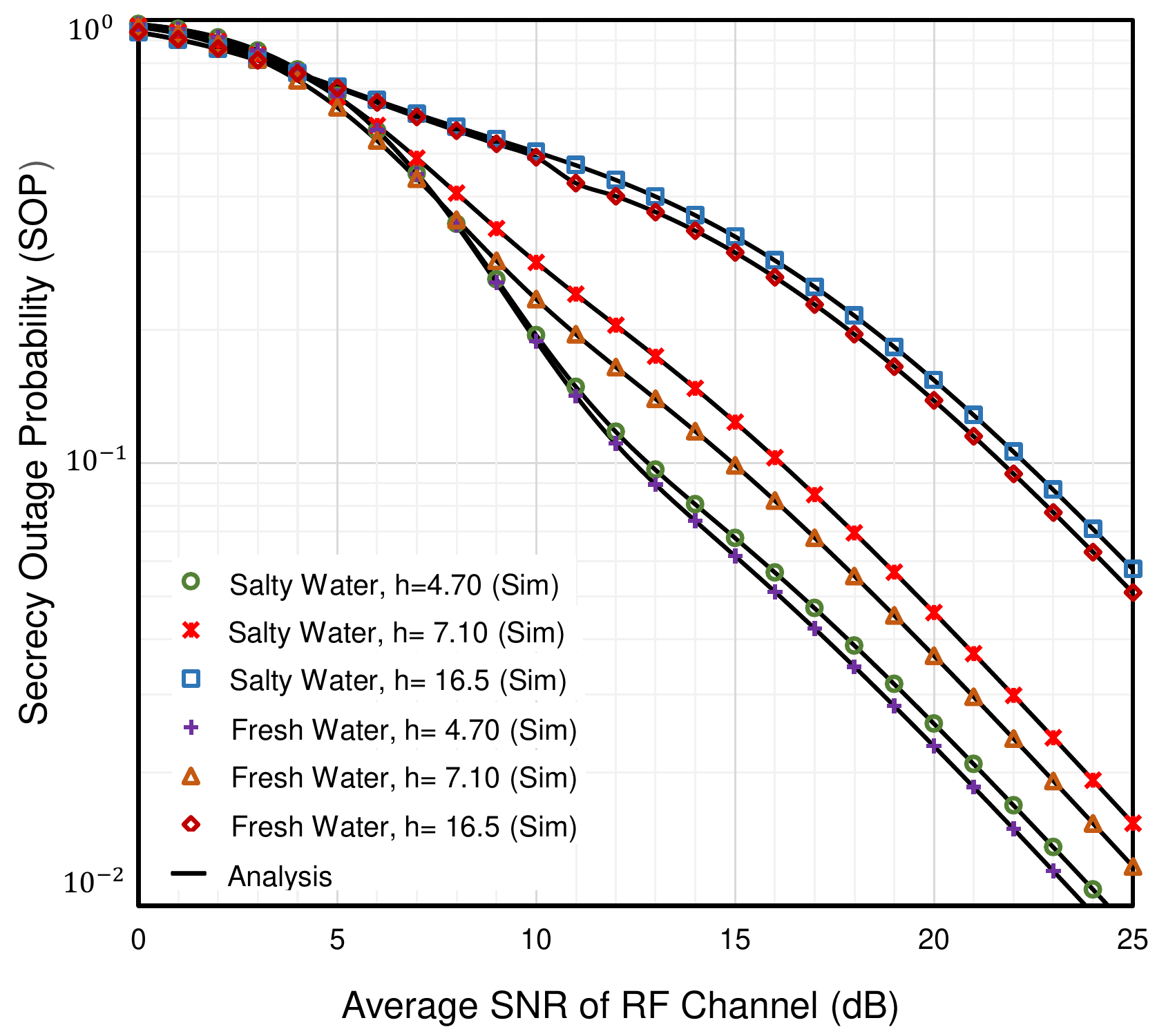}}
        \vspace{0mm}
    \caption{
    The exact SOP versus $\gamma_{m_{1}}$ for specific values of $h$ with $m_{r_{1}}=m_{r_{2}}=m_{e_{1}}=m_{e_{2}}=S_{1}=S_{2}=2$, $\Omega_{r_{1}}=\Omega_{r_{2}}=\Omega_{e_{1}}=\Omega_{e_{2}}=1$, $\gamma_{m_{2}}=0$ dB, $r=1$, $\gamma_{m_{u}}=30$ dB, and $\epsilon_{0}=0.5$ bits/Hz.
    }
    \label{SOPE_h_l_vary}
\end{figure}
\begin{figure}  [!ht]
\vspace{0mm}
    \centerline{\includegraphics[width=0.45\textwidth,angle =0]{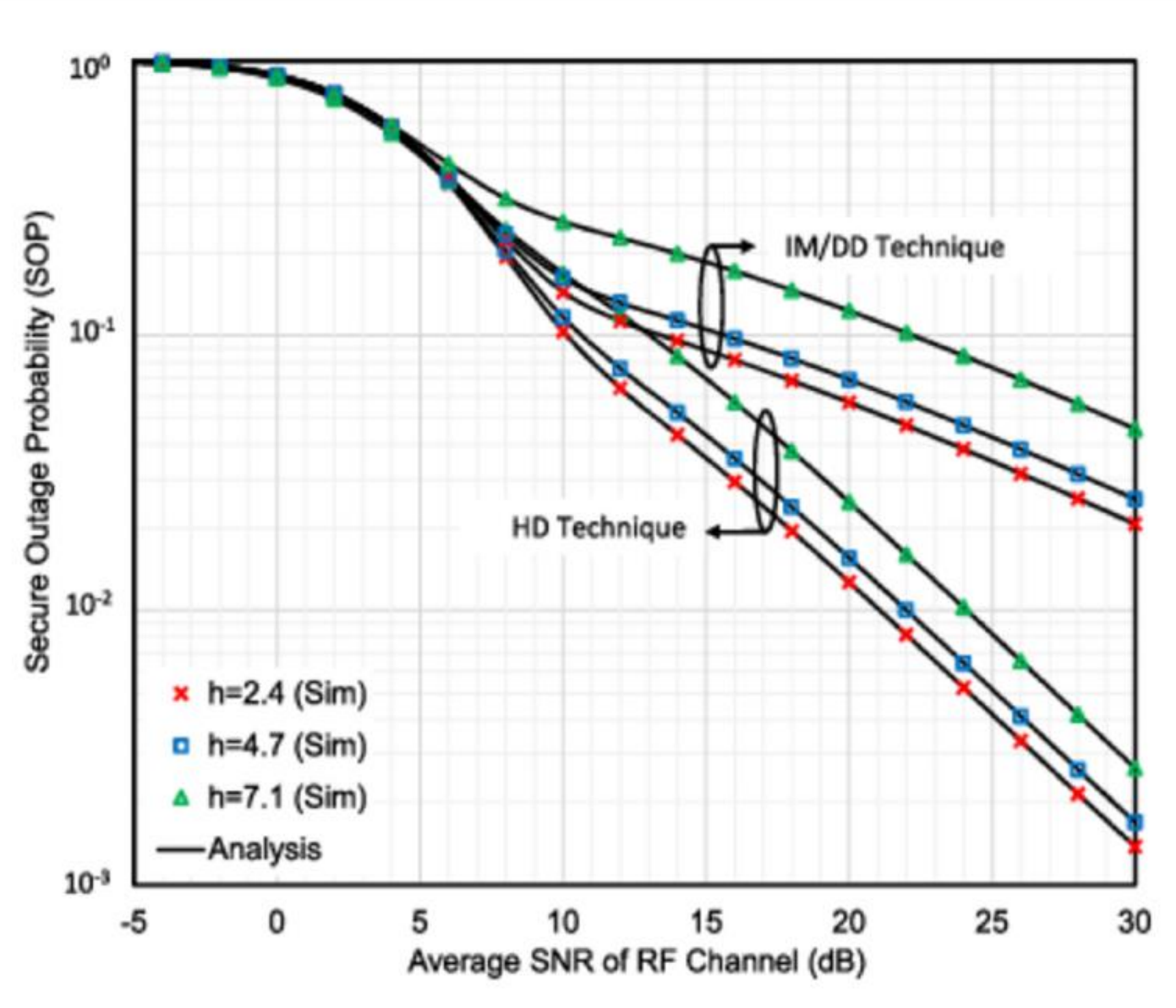}}
        \vspace{0mm}
    \caption{
The lower bound of SOP versus $\gamma_{m_{1}}$ for specific values of $h$ and $r$ considering fresh water, $m_{r_{1}}=m_{r_{2}}=m_{e_{1}}=m_{e_{2}}=S_{1}=S_{1}=2$, $\Omega_{r_{1}}=\Omega_{r_{2}}=\Omega_{e_{1}}=\Omega_{e_{2}}=1$, $\gamma_{m_{2}}=0$ dB, $\gamma_{m_{u}}=20$ dB, and $\epsilon_{0}=0.01$ bits/Hz.}
    \label{SOPL_fresh}
\end{figure} 
\begin{figure}  [!ht]
\vspace{0.00mm}
    \centerline{\includegraphics[width=0.45\textwidth,angle =0]{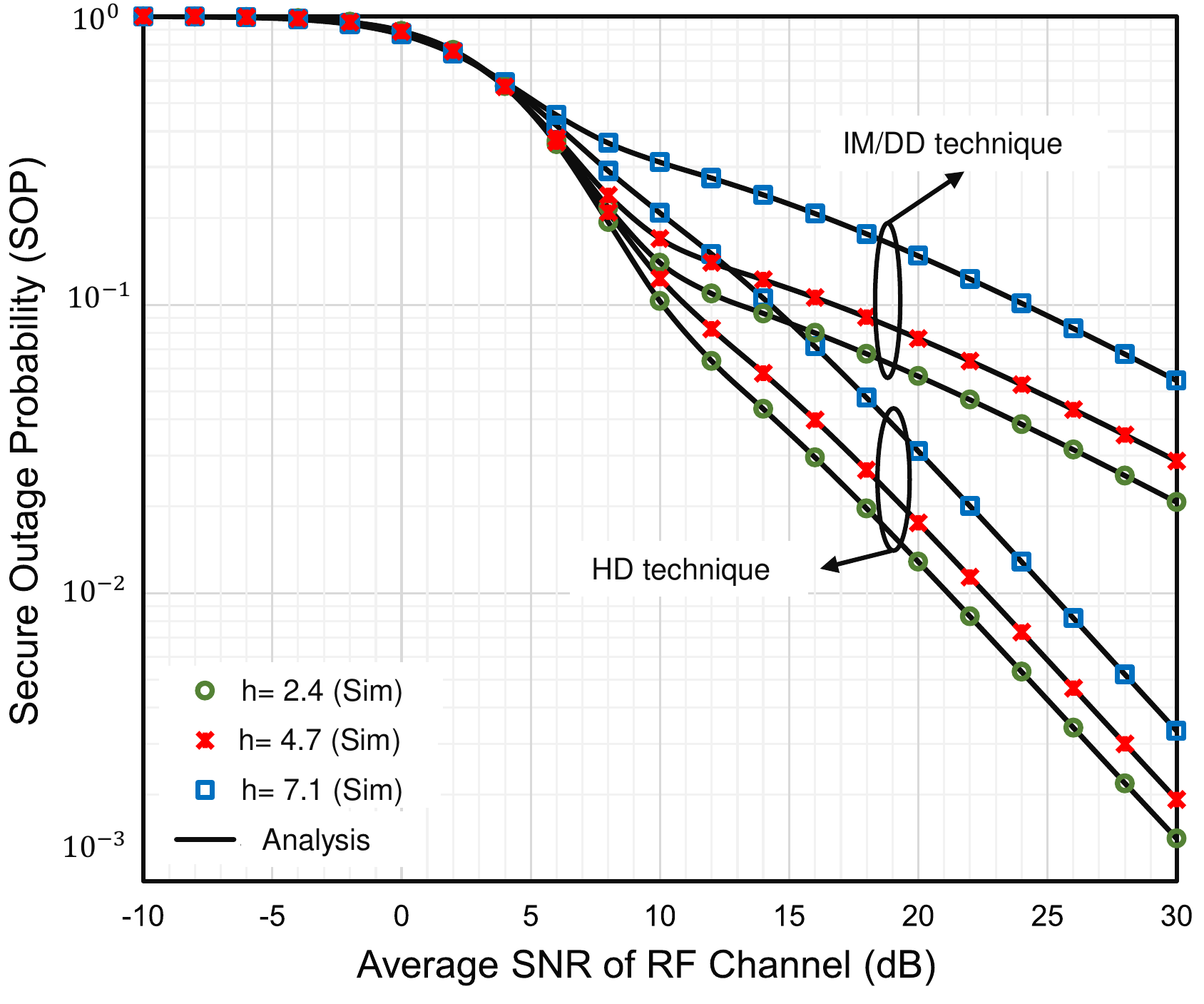}}
        \vspace{0mm}
    \caption{
          The lower bound of SOP versus $\gamma_{m_{1}}$ for selected values of $h$ and $r$ considering salty water, $m_{r_{1}}=m_{r_{2}}=m_{e_{1}}=m_{e_{2}}=S_{1}=2$, $\Omega_{r_{1}}=\Omega_{r_{2}}=\Omega_{e_{1}}=\Omega_{e_{2}}=1$, $\gamma_{m_{2}}=0$ dB, $\gamma_{m_{u}}=20$ dB, and $\epsilon_{0}=0.01$bits/Hz.
    }
    \label{SOPL_Salty}
\end{figure} 
\begin{figure}  [!ht]
\vspace{0.00mm}
    \centerline{\includegraphics[width=0.45\textwidth]{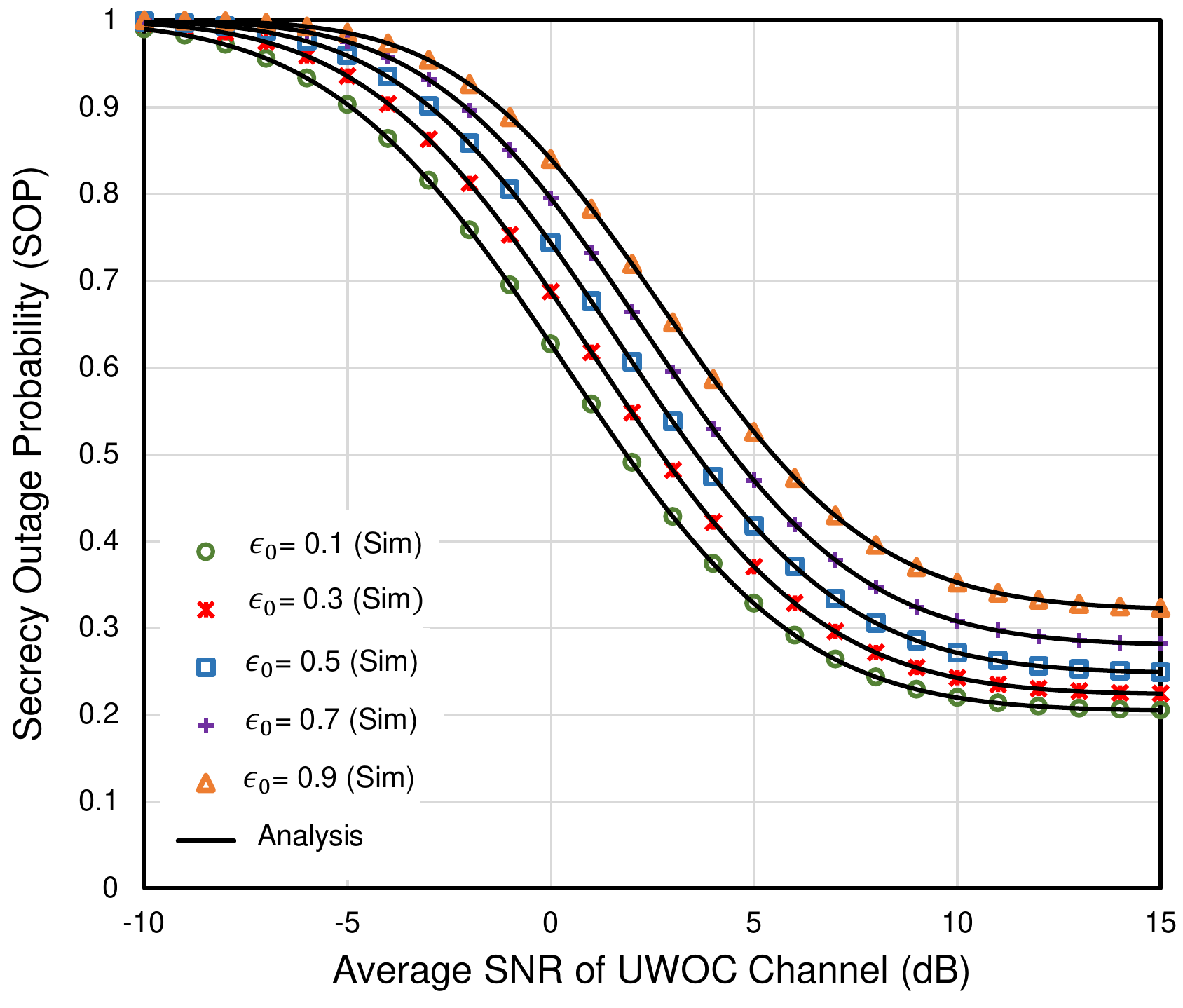}}
    \vspace{0mm}
    \caption{The exact SOP versus $\gamma_{m_{u}}$ for specific values of $\epsilon_{0}$ with $m_{r_{1}}=m_{r_{2}}=m_{e_{1}}=m_{e_{2}}=S_{1}=2$, $\Omega_{r_{1}}=\Omega_{r_{2}}=\Omega_{e_{1}}=\Omega_{e_{2}}=1$,
    $r=2$, $\gamma_{m_{1}}=10$ dB, $\gamma_{m_{2}}=0$ dB, $h=2.4$, and $l=0.05$.}
    \label{SOPE_epsilon}
\end{figure} 
\begin{figure} [!ht]
\vspace{0.00mm}
    \centerline{\includegraphics[width=0.45\textwidth]{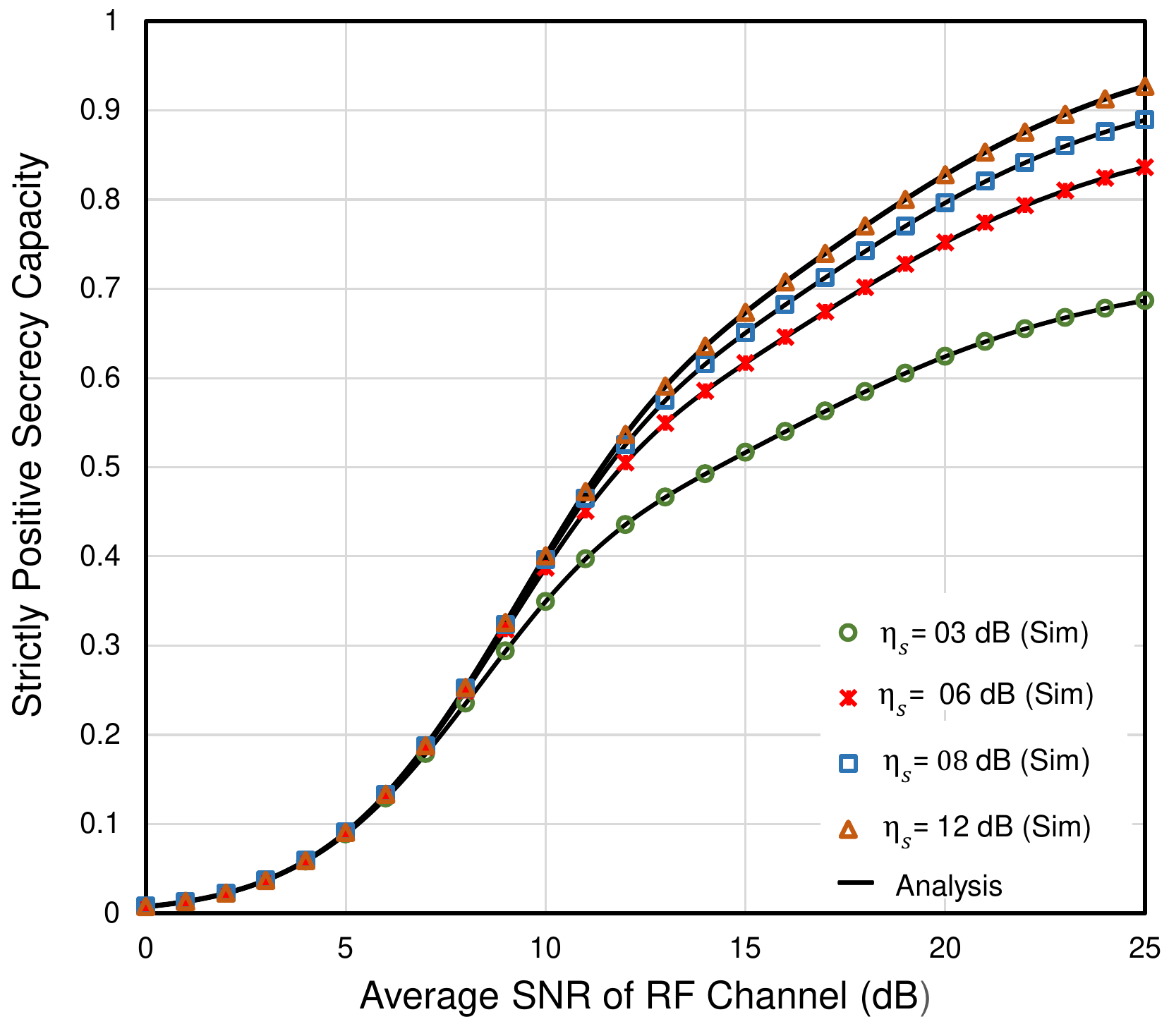}}
    \vspace{0mm}
    \caption{SPSC versus $\gamma_{m_{1}}$ for specific $m_{r_{1}}=m_{r_{2}}=m_{e_{1}}=m_{e_{2}}=S_{1}=S_{2}=2$, $\Omega_{r_{1}}=\Omega_{r_{2}}=\Omega_{e_{1}}=\Omega_{e_{2}}=1$,
    $r=1$, $\epsilon_{0}=0.01$ bits/Hz, $\gamma_{m_{u}}=10$ dB, $\gamma_{m_{2}}=0$ dB, $h=4.7$, and $l=0.1$.}
    \label{SPSC_eta_s}
\end{figure} 
\begin{figure}  [!ht]
\vspace{0.00mm}
    \centerline{\includegraphics[width=0.45\textwidth]{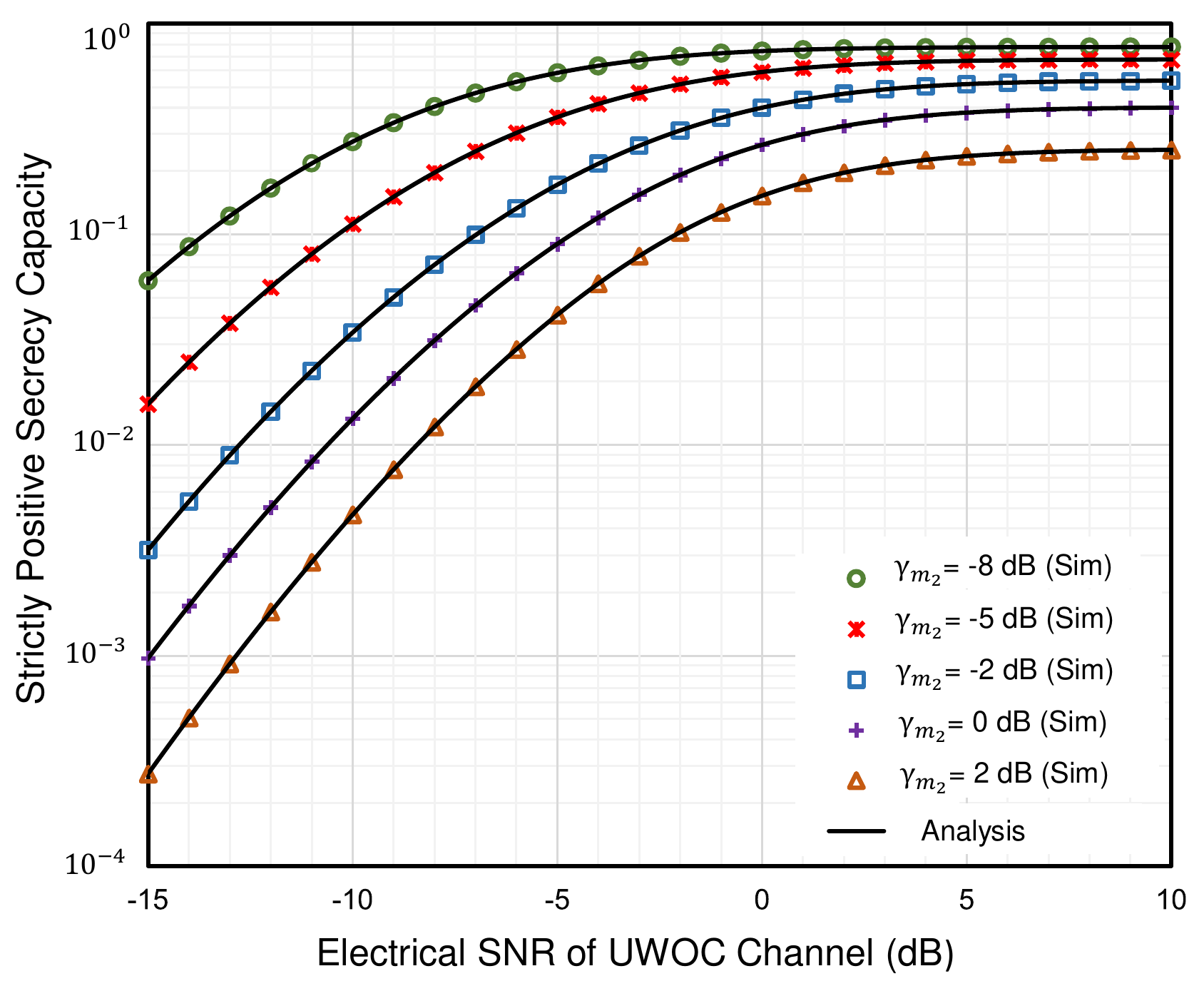}}
    \vspace{0mm}
    \caption{SPSC versus $\eta_{s}$ for specific values of $\gamma_{m_{2}}$ with $m_{r_{1}}=m_{r_{2}}=m_{e_{1}}=m_{e_{2}}=S_{1}=S_{2}=2$, $\Omega_{r_{1}}=\Omega_{r_{2}}=\Omega_{e_{1}}=\Omega_{e_{2}}=1$, $r=1$, $\epsilon_{0}=0.01$ bits/Hz, $\gamma_{m_{1}}=10$ dB, $h=4.7$, and $l=0.1$.}
    \label{SPSC_gamma_m2}
\end{figure}
It is clearly noted that equal increase in $\Omega_{r_{1}}$ and $\Omega_{r_{2}}$, and equal decrease in $\Omega_{e_{1}}$ and $\Omega_{e_{2}}$ enhances the ASC performance. The \textit{Scenario-II} (Fig. \ref{SPSC_omgr2=omge2}) also draws the same conclusion on the impacts of shape parameters as \textit{Scenario-I}, which reveals that the secrecy performance can be improved by increasing $\Omega_{r_{2}}$ and decreasing $\Omega_{e_{2}}$ while $\Omega_{r_{1}}=\Omega_{e_{1}}$ is kept constant.

The impact of varying UWT cases under a thermal gradient system are investigated via plotting SOP against $\gamma_{m_{1}}$ in Fig. \ref{SOPL_therm_grad}.
Note that with the increase in level of the air bubbles and/or temperature gradient, the scintillation index becomes higher causing a stronger UWT and hence the SOP performance is deteriorated. For example, at $\gamma_{m_{1}}=20$ dB, the SOP is $0.09076$ for $h= 2.4$ L/min and the SOP increases to $0.20344$ for $h= 4.7$ L/min with a fixed temperature gradient of $0.05^{0}$ C.$cm^{-1}$. Likewise, for a fixed air bubbles level of $h= 4.7$ L/min, the SOP at $\gamma_{m_{1}}=20$ dB is $0.20344$ for $l=0.05^{0}$ C.$cm^{-1}$ whereas the SOP increases to $0.23833$ for $l=0.10^{0}$ C.$cm^{-1}$. It is also observed that the impact of the temperature gradient is more significant than the air bubbles level since the temperature gradient can induce stronger irradiance fluctuations leading to a severe UWT. Same conclusions are also drawn in \cite{badrudduza2021security} that proves the accuracy of our results.

The impact of UWT under a thermally uniform UOWC system is presented in Fig. \ref{SOPE_h_l_vary} for both fresh and salty waters.
Note that, similar to Fig. \ref{SOPL_therm_grad}, similar effects of air bubbles level are observed for both types of waters. It is also observed that additional salinity increases the UWT but here the impact of air bubbles in inducing UWT is more significant. Hence, the secure outage performance of fresh water is better than that of salty water environment as testified in \cite{badrudduza2021security}.

Figs. \ref{SOPL_fresh} and \ref{SOPL_Salty} depict a comparison between HD and IM/DD techniques in terms of SOP performance.
It is noteworthy that HD technique overcomes UWTs more remarkably relative to IM/DD technique leading to a significant improvement in the secrecy performance. This is because the HD technique includes implementation of coherent receivers that may arise some complexity in the structures but definitely a better SNR is obtained at the destination relative to the IM/DD technique \cite{badrudduza2021security}.

In Fig. \ref{SOPE_epsilon}, SOP is investigated against $\gamma_{m_{u}}$ to observe the impact of $\epsilon_{0}$.
It is clearly observed similar to \cite{mandira2021secrecy} that outage performance degrades with $\epsilon_{0}$. When $\epsilon_{0}$ is considered to be a high value, $C_{s}$ starts to fall below $\epsilon_{0}$ that in turn results in the degradation of system outage performance. Thus, to ensure a secure communication over RIS-aided UOWC link, $C_{s}$ must be higher than $\epsilon_{0}$.


Fig. \ref{SPSC_eta_s} illustrates SPSC performance under the effect of $\eta_{s}$ considering HD technique (i.e. $s=1$).
Higher values of $\eta_{s}$ improves the $\mathcal{T}-\mathcal{I}_{R}-\mathcal{R}$ link quality, as a result, the SPSC performance is also enhanced with the increase in $\eta_{s}$ but the system performance reorients at lower values of $\gamma_{{m}_{1}}$ and stays steady in the SPSC ceiling of each curve.

To evaluate the effects of $\gamma_{m_{2}}$, Fig. \ref{SPSC_gamma_m2} demonstrates that SPSC performance deteriorates as soon as $\gamma_{m_{2}}$ changes from weaker ($-8$ dB)-to-stronger ($2$ dB) conditions.
Clearly, a higher $\gamma_{m_{2}}$ guarantees stronger eavesdropper channel causing the SPSC to degrade. It further indicates that at higher values of the electrical SNR of the UOWC channel, the SPSC remains almost unchanged and a ceiling is observed with a little difference among the plotted five curves for $\gamma_{m_{2}}=-8$ dB, $-5$ dB, $-2$ dB, $0$ dB, and $2$ dB. This is because the secrecy performance of the proposed RIS-aided model is always dominated by the weaker hop \cite{samuh2020performance}.

\section{Conclusions}
In this research, we inspect the secrecy behavior of a combined dual-hop RIS-assisted RF-UOWC framework. To understand the insights of this scheme, mathematical expressions of performance measures (i.e. ASC, SOP, and SPSC) are derived appraising the generalized system properties of the RF and UOWC channels. Our derived expressions are also validated by the emulation of the MC simulations. The outcomes demonstrate that the number of reflecting elements facilitates better secrecy performance by increasing SNR gain and HD technique outperforms the IM/DD technique to ensure the perfect secrecy level. Secure communication is assured only if the minimum number of reflecting elements of RIS for both user and eavesdropper is ensured. Furthermore, the influence of fading parameters, scale parameters, air bubbles levels, temperature gradients, and water salinity are inspected. The expansion of this work can be stretched considering multiple colluding and non-colluding eavesdroppers in both surface RF and underwater scenarios.

\bibliographystyle{IEEEtran}
\bibliography{IEEEabrv,asmbBiblio.bib}

\end{document}